\documentclass[twocolumn,epjc3]{svjour3mod}  
\journalname{Eur. Phys. J. C}
\usepackage[utf8]{inputenc}
\usepackage{epsfig,amsfonts,amssymb,latexsym,amsmath,bm,color,array,epsfig,graphics,dsfont}
\usepackage{enumitem} 
\usepackage{amsmath}
\usepackage{amssymb}
\usepackage{physics} 
\usepackage{lastpage}
\usepackage{graphicx}
\usepackage{xcolor}
\usepackage[colorlinks,citecolor=blue,urlcolor=blue,linkcolor=blue,breaklinks]{hyperref}
\usepackage{balance,cite}
\usepackage{booktabs}
\usepackage{multirow}
\usepackage{subcaption}

\newcommand{\mpi}{M_\pi}
\newcommand{\beq}{\begin{equation}}
\newcommand{\eeq}{\end{equation}}
\newcommand{\diff}{\text{d}}

\newcommand{\BR}{\text{Br}}

\newcommand{\res}{\textrm{R}}
\newcommand{\bg}{\textrm{B}}
\newcommand{\sh}{\text{sh}}
\newcommand{\thr}{\text{thr}}

\newcommand{\GeV}{\,\text{GeV}}
\newcommand{\MeV}{\,\text{MeV}}

\newcommand{\Mcal}{\mathcal{M}}
\newcommand{\Acal}{\mathcal{A}}
\newcommand{\Order}{\mathcal{O}}

\allowdisplaybreaks[1]

\numberwithin{equation}{section}

\begin{document}
\emergencystretch 3em

\title{From pole parameters to line shapes and branching ratios}

\author{L.~A.~Heuser\thanksref{add1,add4,e1} \and \nolinebreak[4]
G.~Chanturia\thanksref{add4,e2} \and \nolinebreak[4]
\mbox{F.-K.~Guo}\thanksref{add2,add5,add6} \and \nolinebreak[4]
C.~Hanhart\thanksref{add1} \and \nolinebreak[4]
M.~Hoferichter\thanksref{add3} \and \nolinebreak[4]
B.~Kubis\thanksref{add4}}

\thankstext{e1}{e-mail:~heuser@hiskp.uni-bonn.de}
\thankstext{e2}{e-mail:~g.chanturia@uni-bonn.de}
%
 
\institute
{Forschungszentrum J\"ulich, Institute for Advanced Simulation,  52425 J\"ulich, Germany \label{add1}
\and
Helmholtz-Institut f\"ur Strahlen- und Kernphysik and Bethe Center for Theoretical Physics, Universit\"at Bonn, 53115 Bonn, Germany \label{add4}
\and 
CAS Key Laboratory of Theoretical Physics, Institute of Theoretical Physics, Chinese Academy of Sciences, Beijing 100190, China \label{add2}
\and
School of Physical Sciences, University of Chinese Academy of Sciences, Beijing 100049, China \label{add5}
\and
Peng Huanwu Collaborative Center for Research and Education, Beihang University, Beijing 100191, China \label{add6}
\and
Albert Einstein Center for Fundamental Physics, Institute for Theoretical Physics, University of Bern, Sidlerstrasse 5, 3012 Bern, Switzerland \label{add3}
}
\date{}

\maketitle

\begin{abstract}
Resonances are uniquely characterized by their complex pole locations and the 
corresponding residues. In practice, however, resonances are typically identified experimentally as structures in invariant mass distributions, with branching fractions of resonances determined as ratios of count rates. To make contact between these quantities it is necessary to connect line shapes and resonance parameters. 
In this work we propose such a connection and illustrate the formalism
with detailed studies of the $\rho(770)$ and $f_0(500)$ resonances.  
Based on the line shapes inferred from the resonance parameters along these lines, expressions for partial widths and branching ratios are derived
and compared to other approaches
in the literature.
\end{abstract}

\section{Introduction}
\label{sec:intro}

Most hadronic states are not stable in quantum chromodynamics (QCD) and possess a decay width too large to be approximated by a pole on the real axis. Instead, such resonances are described mathematically by poles in the complex-energy plane, and their characterization therefore requires an analytic continuation of the scattering matrix. In this way,  
the coupling of a resonance to a given decay
channel is determined by the residue, that is, the strength of
the scattering amplitude at the resonance pole. In practice, however, this analytic continuation can be highly non-trivial, and the projection onto the real axis, where experiments are performed, can differ widely depending on the complexity of the system. This ranges from clear-cut cases such as the $\rho(770)$, in which the resonance peak is clearly visible in the cross section, via examples such as the $f_0(500)$, in which case only a broad bump is observed, to complicated multi-channel systems such as the $f_0(980)$, which may show up as a narrow peak or a dip structure. In the last example, the line shape can differ so dramatically depending on the source that drives its production, since this is what
controls the interference pattern of the given resonance with its background. 
It is therefore not at all straightforward
to experimentally define branching ratios for a resonance. 
Even in cases in which the resonance pole parameters can be determined reliably via dispersive analyses of the scattering matrix, 
the concept of branching ratios needs to be understood theoretically in terms of pole 
parameters. The main goal of this work is to establish such a connection.

The framework we propose here is based on the two-potential formalism~\cite{Nakano:1982bc}, constructed in such a way that constraints from analyticity and unitarity are maintained, while allowing for enough freedom to parameterize the effects of left-hand cuts (LHCs). To argue that such an approach constitutes, in fact, a minimal solution to the general problem, we proceed as follows. After defining the formalism in Sec.~\ref{sec:Swaves}  for $S$-waves and discussing its generalization to higher partial waves in Sec.~\ref{sec:higherpw}, with some details on conventions relegated to~\ref{app:conventions} and~\ref{app:radiative}, we start in Sec.~\ref{sec:rho770} with the application to the $\rho(770)$, a resonance structure so clear that even the Breit--Wigner ansatz~\cite{Breit:1936zzb} gives a reasonable description. As a first step to improve beyond such a model, self-energy corrections need to be included to restore the correct analyticity properties, which leads to a form closely resembling the Gounaris--Sakurai parameterization of the $\rho(770)$~\cite{Gounaris:1968mw}. However, we will show that with this procedure only real and imaginary parts of the pole location,
\begin{equation}
\sqrt{s_\res} = M_\res-i\frac{\Gamma_\res}{2}\, ,
\label{eqn:MGamdef}
\end{equation}
by convention expressed in terms of the pole mass $M_\res$ and pole width $\Gamma_\res$, can be reproduced exactly, while the residue is predicted in terms of these parameters. For a high-precision description of pion--pion ($\pi\pi$) scattering~\cite{Ananthanarayan:2000ht,Colangelo:2001df,Garcia-Martin:2011iqs,Caprini:2011ky,Colangelo:2018mtw} and the resulting $\rho(770)$ pole parameters obtained from analytic continuation of the Roy equations~\cite{Roy:1971tc}, this does not provide sufficient flexibility. Reproducing the $\rho(770)$ parameters at the precision level is not only important to illustrate how our formalism works, but also of phenomenological interest, as a starting point to describe $4\pi$ inelasticities in the electromagnetic form factor of the pion~\cite{Hanhart:2012wi,Chanturia:2022rcz}, which is critical for a better understanding of tensions in the $2\pi$ contribution to hadronic vacuum polarization~\cite{Colangelo:2018mtw,Colangelo:2020lcg,Colangelo:2022prz}.

Next in complexity we turn to the $f_0(500)$ in Sec.~\ref{sec:f0500}. While the existence of this lowest-lying resonance in QCD was contested for decades~\cite{Pelaez:2015qba}, the required analytic continuation deep into the complex plane can again be performed in a reliable manner based on dispersion relations~\cite{Caprini:2005zr,Garcia-Martin:2011nna,Moussallam:2011zg}, despite the fact that the $f_0(500)$ is not visible in the $\pi\pi$ $S$-wave phase shift as a clear resonance structure (the same is true for scalar form factors, see, e.g., Refs.~\cite{Donoghue:1990xh,Gardner:2001gc,Hoferichter:2012wf,Daub:2015xja}). Moreover, in this case the presence of an Adler zero~\cite{Adler:1964um,Adler:1965ga} is critical to obtain a realistic line shape. For instance, it is known from the inverse-amplitude method~\cite{Dobado:1996ps,GomezNicola:2007qj} that unitarizing amplitudes from chiral perturbation theory (ChPT) with the right Adler zero, the $f_0(500)$ parameters are reproduced with reasonable accuracy. Here, we will show the opposite direction, finding that starting from the $f_0(500)$ resonance parameters, our formalism automatically produces an Adler zero in the vicinity of its ChPT expectation. We also detail how the correct threshold behavior of the LHCs can be incorporated, see~\ref{app:LHC}, and evaluate higher-order chiral corrections to the Adler zero, see~\ref{app:Adler}.   

Having demonstrated how our formalism recovers the $\rho(770)$ and $f_0(500)$ as resonances in $\pi\pi$ scattering, we turn to the generalization to multi-channel systems in Sec.~\ref{sec:branchings}. In such a case, if a resonance couples to various channels, the imaginary part of the pole position
acquires contributions from all of them. Depending on the Riemann sheet on which
the most significant pole is located, the individual imaginary parts not necessarily
add, and some care is required in defining consistent branching fractions and decay widths, see, e.g., Refs.~\cite{Wang:2022vga,Svarc:2022buh,Burkert:2022bqo} for recent works in this direction. While usually the problem is phrased as the determination of pole parameters from the analytic continuation of scattering amplitudes~\cite{Battaglieri:2014gca}, we take here the opposite perspective and discuss 
 to what extent line shapes, and from those branching ratios
of resonances, can be deduced from a set of pole parameters. The main goal is to replace common prescriptions to turn residues into branching fractions by a better justified recipe. For example, a narrow-width formula for the $f_0(500)\to\gamma\gamma$ decay~\cite{Morgan:1987gv,Morgan:1990kw,Hoferichter:2011wk,Moussallam:2011zg} fails to account for the complicated line shape of the $f_0(500)$, while the branching ratio for $f_0(500)\to\bar K K$~\cite{Danilkin:2020pak} would even vanish, since the resonance mass lies below the $\bar K K$ threshold. Instead, we will show how our formalism allows us to derive well-defined, normalized spectral functions, from which partial widths and branching ratios can be inferred in a consistent manner. As test cases, we again consider $\rho(770)$ and $f_0(500)$, comparing our prescription to other proposals in the literature.  Our formalism can be generalized to more complicated cases such as the $f_0(980)$~\cite{Zou:1993az,Garcia-Martin:2011nna,Moussallam:2011zg} or $a_0(980)$~\cite{Albaladejo:2015aca,Lu:2020qeo}, for which different Riemann sheets play a role.  In Sec.~\ref{sec:summary} we summarize our main results and give an outlook towards such future applications.

\section{$S$-wave formalism}
\label{sec:Swaves}

\subsection{Scattering amplitude and residues}

All information on a scattering process is encoded in the scattering amplitude $\cal M$, connected to the $S$-matrix via
\begin{align} 
& _{\text{out}}\!\bra{p_{1}' p_{2}',b} \mathcal{S} - 1 \ket{p_1 p_2,a}_\text{in}\nonumber & \\
&\qquad =  i(2\pi)^4\delta^4(p_1+p_2-p_{1}'-p_{2}')\,\Mcal_{ba} \, ,
\end{align}
where, for concreteness, we concentrate on a two-to-two reaction.
Close to the resonance pole it can be expanded into a Laurent series as 
\begin{equation}
\Mcal_{ba} = - \frac{\mathcal{R}_{ba}}{s-s_\res} + \mbox{regular terms} \, ,
\label{eqn:resdef}
\end{equation}
where $a$ and $b$ are channel indices.
The residue $\mathcal{R}_{ba}$ can be conveniently extracted from the amplitude via 
\begin{equation}
\mathcal{R}_{ba} = -\frac{1}{2\pi i}\oint ds \, \Mcal_{ba} \, ,
\end{equation}
where the closed integration path needs to be chosen such that
it runs counterclockwise and the pole of interest is the
only non-analyticity enclosed.
The factorization of the residue $(\mathcal{R}_{ba})^2 = \mathcal{R}_{aa}\times \mathcal{R}_{bb}$
allows one to introduce pole couplings according to
\begin{equation}
\tilde g_a = H(s_p)\mathcal{R}_{ba} /\sqrt{\mathcal{R}_{bb}}  \, .
\label{eqn:gpoledef}
\end{equation}
The function $H(s_p)$
is introduced here to collect convention-dependent factors often 
introduced for the effective couplings, e.g., for higher partial waves $H(s_p)$ traditionally absorbs the
centrifugal barrier factor. The conventions relevant for the effective couplings employed in this work are provided in~\ref{app:conventions}.
It should be stressed
that these pole couplings are the only model- and reaction-independent quantities that allow one 
to quantify the
transition strength of a given resonance to some channel $a$.

\subsection{Dyson series and self energy}

As a starting point, we consider
 the case of a resonance coupling to a single continuum channel in an $S$-wave.
Higher partial waves are discussed in Sec.~\ref{sec:higherpw} and
   the
generalization to  more channels is provided in 
 Sec.~\ref{sec:branchings}, where also partial widths and branching ratios are introduced.
Theoretically, the physical propagator of a single resonance, $G(s)$,  emerges as the solution
of the Dyson equation for some given self-energy function $\Sigma(s)$:\footnote{We define the self energy without the coupling,
as this allows us to keep track of the parameters appearing in the formalism that are independent of the dynamics.} 
\begin{equation}
G(s) = G_0(s)-G_0(s)g^2\Sigma(s)G(s)\, ,
\label{eq:Dyson}
\end{equation}
with the bare propagator
\begin{equation}
G_0(s)=(s-m^2)^{-1}\, .
\end{equation}
Equation~\eqref{eq:Dyson} is solved by
\begin{equation}
G(s) = \big(s-m^2+g^2\Sigma(s)\big)^{-1} \, .
\label{Gsimple}
\end{equation}
Unitarity requires both $g$ and $m$ to be real parameters. The
self energy $\Sigma(s)$ contains all one-particle irreducible diagrams with respect to the studied resonance that contribute
to the two-point function in the resonance channel.

In the simplest scenario in which there is no background term and the complete interaction of the scattering particles is provided by the resonance one has
\begin{equation}
   \text{disc}\,\Sigma(s) = 2i\rho(s) \, ,
   \label{eq:discsimple}
\end{equation}
where 
\begin{equation}
   \rho(s)=\frac1{16\pi}\frac{2q}{\sqrt{s}}\, , \qquad q=\frac12\sqrt{s-4M^2} \, ,
\end{equation}
$M$ is the mass of the particles in the continuum channel, and $q$ denotes the momentum of the outgoing particles in the center-of-mass frame. In this work we 
mostly study channels with particles of equal mass, however, the generalization to different
masses is straightforward.
In case of absence of a background term, such that the
discontinuity is provided by
Eq.~\eqref{eq:discsimple},
the self energy  $\Sigma(s)$ equals the polarization function $\Pi(s)$, which can
be written as a once-subtracted dispersion integral
\begin{equation}
   \Pi(s) = b + \frac{s-s_0}{\pi}\int_{s_\thr}^\infty \frac{\diff s'}{s'-s_0}\frac{\rho(s')}{s'-s} = b + \Pi^r(s)  \, ,
   \label{Pidef}
\end{equation}
with some subtraction constant $b$ that can be absorbed into other parameters of the amplitude. The scattering threshold $4M^2$ is denoted as $s_\thr$, and $s_0$ is the subtraction point. The index $r$ indicates
that $\Pi^r(s)$ is the renormalized self energy. Since from now on all self energies are renormalized,
we drop the index $r$ again to ease notation.
For $s_0=4M^2$ one finds 
\begin{equation}
   \Pi(s)=\frac{\rho(s)}{\pi}\log\left(\frac{16\pi\rho(s)-1}{16\pi\rho(s)+1}\right) 
 \end{equation}
for all values of $s$ on the first sheet.
Under these conditions the
scattering amplitude reads
\begin{equation}
   \Mcal(s)=-\frac{g^2}{s-m^2+g^2\Pi(s)} \, .
   \label{Tsimple}
\end{equation}
To obtain the correct resonance pole location of $\Mcal$, one therefore has to demand
\begin{align}
   \Im s_\res&= -g^2\Im\left(\Pi^{(-)}(s_\res)\right) \, , \notag \\
   \Re s_\res&= m^2-g^2 \Re\left(\Pi^{(-)}(s_\res)\right) \, ,
   \label{fixinggm}
 \end{align}
where the superindex $(-)$ indicates that the pole location of a resonance
is on the unphysical sheet that is defined by $\Im q <0$.
However, by imposing the conditions of Eq.~\eqref{fixinggm} the scattering amplitude of
Eq.~\eqref{Tsimple} is fixed completely.
In particular we then find for the effective coupling (setting for simplicity $H(s_p)$ from Eq.~\eqref{eqn:gpoledef} to $1$ for the $S$-wave case discussed here)
\begin{equation}
   \tilde g^2 =  Zg^2\, ,\qquad Z=\bigg(1+g^2\frac{\diff\Pi^{(-)}(s)}{\diff s}\bigg|_{s=s_\res}\bigg)^{-1} \, .
\label{Zdef}
\end{equation}
In some cases this already allows for a fair representation of the pole parameters; in fact, the $P$-wave version of
Eq.~\eqref{Tsimple} closely resembles  the venerable Gounaris--Sakurai
parameterization for the $\rho(770)$~\cite{Gounaris:1968mw}.
However, Eq.~\eqref{Tsimple} does
not have sufficient flexibility to fix pole location and residue independently, which becomes problematic for a precision description of the $\rho(770)$, and, as we will demonstrate below, it fails badly
for the scalar--isoscalar $\pi\pi$ $S$-wave.

\subsection{Two-potential formalism}

The goal of this work is
to find a more general expression for the resonance propagator
that is consistent with the fundamental field theoretic principles of
unitarity, analyticity, and positivity of the spectral function of the full propagator.  
To reach this goal 
we employ the two-potential formalism~\cite{Nakano:1982bc}.
It allows one to decompose the full scattering amplitude as 
\begin{equation}
   \Mcal(s) = \Mcal_\bg(s) + \Mcal_\res(s) \, ,
   \label{Ttotdef}
\end{equation}
where $\Mcal_\bg(s)$ denotes some properly chosen background amplitude. For example,
in Refs.~\cite{Hanhart:2012wi, Ropertz:2018stk} $\Mcal_\bg(s)$
was chosen in such a way that the full scattering amplitude at low energies reproduced the
high-precision $\pi\pi$ phase shifts from Refs.~\cite{Ananthanarayan:2000ht,Colangelo:2001df,Garcia-Martin:2011iqs}, and similarly for $\pi K$ scattering in Ref.~\cite{VonDetten:2021rax}.
In this way it is possible to import pertinent information on the LHCs into the resonance formalism. On the other hand, it does not
allow for a straightforward evaluation of the amplitude at the resonance pole,
since a continuation to the second sheet calls for
an analytic continuation of the input scattering amplitude $\Mcal_\bg$, which is not known in this case, cf.~Eq.~\eqref{gam+sigcont} below.
Therefore, we here employ  some explicit
representation of the background term that allows us
to perform the mentioned analytic continuation.

Since the full scattering amplitude respects the unitarity relation and so does $\Mcal_\bg$, this does not hold for $\Mcal_\res$ by itself. In particular one finds
\begin{align}
  \Mcal_\res(s)&=  -  \frac{g^2\gamma^2(s)}{s-m^2+g^2\Sigma(s)}\notag\\&
  \equiv
  - \gamma(s)g\, G_\res(s)\, g\gamma(s)
  \label{TRdef}
\end{align}
for the resonance part of the scattering amplitude, with the self energy $\Sigma(s)$ now dressed by the vertex function $\gamma(s)$ to be constructed below, and
\begin{equation}
  \Acal_\res(s)=  -\gamma(s)g  \, G_\res(s) \, \alpha 
  \label{FFdef}
\end{equation}
for the production amplitude  (up to a multiplicative polynomial)
that originates from the resonance, with
$\alpha$ quantifying the resonance--source coupling. Equation~\eqref{TRdef} defines the physical resonance propagator $G_\res(s)$.
On the physical axis the vertex function $\gamma(s)$ and the dressed self energy $\Sigma(s)$ are now linked to the background amplitude via
\begin{align} 
 \text{disc}\,\gamma(s) &=  2i\rho(s) \Mcal_\bg(s)^*\gamma(s) \, , \notag \\
   \text{disc}\,\Sigma(s) &= 2i\rho(s)\left|\gamma(s)\right|^2 \, .
   \label{discgam+sig}
\end{align}
In this way the particle pairs propagating from the vertex
or within
the loop are not moving freely (as they do in Eq.~\eqref{Pidef}),
but  undergo interactions driven by $\Mcal_\bg(s)$.
Equation~\eqref{discgam+sig} 
at the same time provides a prescription for the analytic
continuation of both vertex function and self energy into the unphysical sheet of the complex $s$ plane, via
\begin{align} 
 \gamma^{(-)}(s) &=  \gamma(s)\left(1 - 2i\rho(s) \Mcal_\bg^{(-)}(s)\right) \, , \notag\\
   \Sigma^{(-)}(s) &= \Sigma(s) - 2i\rho(s)\gamma^{(-)}(s)\gamma(s) \, ,
   \label{gam+sigcont}
\end{align}
where we need to use $\rho(s^*)=-\rho(s)^*$ for the analytic continuation of
the phase-space factor from the upper to the lower half of the complex $s$ plane~\cite{Pelaez:2015qba}.

\subsection{Explicit parameterizations}

To allow for an analytic continuation of $\Mcal_\bg$ needed in Eq.~\eqref{gam+sigcont}, 
we employ an explicit  parameterization:
\begin{equation}
   \Mcal_\bg(s)=\frac{f_0}{f(s)-f_0\Pi(s)} \equiv \frac{1}{\rho(s)}\sin\delta_\bg(s) e^{i\delta_\bg(s)} \, ,
   \label{Tbdef}
\end{equation}
where the background phase $\delta_{\rm B}$ in the expression on the far right is defined for real values of $s$ above the scattering threshold only. 
For $\Mcal_\bg\equiv 0$ (achieved by $f_0\to 0$)
we recover the simple scattering amplitude provided in Eq.~\eqref{Tsimple}.
In the general case, however, the parameter $f_0$ and
the function $f(s)$ allow us to vary both strength
and phase of the residue independently of the pole location.
Moreover, we can even effectively include LHCs into $\Mcal_\bg$
 by employing a polynomial in a 
properly chosen conformal variable $\omega(s)$~\cite{Gasparyan:2012km,Pelaez:2019eqa,Danilkin:2022cnj}:
\begin{equation}
   f(s)=1+\sum_{k=1}^{k_{\rm max}} f_k \omega^k(s) + f_R s \, .
   \label{fofsdef}
\end{equation}
The parameter $f_R$ is introduced to ensure that $\lim_{s\to \infty}\Mcal_\bg(s)=0$,
such that $\Mcal_\bg$ and with it also $\Mcal_\res$ drop as $1/s$ for large values of 
$s$. It is not employed in
the fit to the residues but is kept fixed at some sufficiently small value to
keep its effect small in the resonance region. For example, in the study of the $\rho(770)$ and $f_0(500)$ presented below, we use $f_R = 1/(2\GeV)^2$, including the variation to $f_R=1/(3\GeV)^2$ in the final uncertainty estimates. 
It should be stressed that there is no guarantee that the given parameterization
for $f(s)$ does not lead to unphysical poles, so that checking for their absence
is to be part of the analysis.
For $\omega(s)$ we use the
prescription~\cite{Danilkin:2022cnj}
\begin{equation}
   \omega(s) = \frac{\sqrt{s-s_L}-\sqrt{s_E-s_L}}{\sqrt{s-s_L}+\sqrt{s_E-s_L}} \, ,
\end{equation}
where $s_L$ denotes the location of the closest branch point of the LHC---for
$\pi\pi$ scattering one has $s_L=0$---and $s_E$ some conveniently
chosen expansion point; we use $s_E=\Re s_\res$. 
In the case of $\pi\pi$ scattering the leading
LHC arises from
two-pion exchange in the $t$- and $u$-channel, whose partial wave projection
for both $\pi\pi$ $S$- and $P$-waves leads to an onset
of the LHC scaling as $(-s)^{3/2}$ near $s=0$; see \ref{app:LHC}.
To implement this property, instead of using the parameter
$\omega$ directly in Eq.~\eqref{fofsdef}
we expand in $2\omega(s)+[\omega(s)]^2$.
Given this parameterization, the analytic continuation of $\Mcal_\bg$ 
 to the unphysical sheet simply goes by replacing $\Pi(s)$ by
$\Pi^{(-)}(s)$ in Eq.~\eqref{Tbdef}, where the latter is given by the
analog of Eq.~\eqref{gam+sigcont} in the absence of a background term, $\gamma(s)\to 1$.

While $\Mcal_\bg$ is allowed to have LHCs, this is not the case for
 $\Mcal_\res$, defined in 
Eq.~\eqref{TRdef}, and the production amplitude $\Acal_\res(s)$, defined in 
Eq.~\eqref{FFdef}. This property is guaranteed by constructing the vertex function
$\gamma(s)$ from the dispersion integral 
\begin{equation}
   \gamma(s)=\exp\left(\frac{s}{\pi}\int_{s_{\thr}}^\infty \frac{\diff s'}{s'}\frac{\delta_\bg(s')}{s'-s}\right) \, ,
   \label{gammaomnes}
\end{equation}
which is the usual once-subtracted Omn\`es function~\cite{Omnes:1958hv}. 
The corresponding subtraction constant is absorbed into the coupling $g$.
It is consistent with the discontinuity equation~\eqref{discgam+sig} and has only the right-hand cut. At the same time the
information on the LHC is imported into $\Mcal_\res(s)$ as well as $\Acal_\res(s)$
via $\delta_\bg(s)$. Analogously, we employ
\begin{equation}
   \Sigma(s) = \frac{s-s_0}{\pi}\int_{s_\thr}^\infty \frac{\diff s'}{s'-s_0}\frac{\rho(s')|\gamma(s')|^2}{s'-s}  
   \label{Sigmadef}
\end{equation}
as a straightforward generalization of Eq.~\eqref{Pidef} in the presence of a background interaction.
In the applications below we choose $s_0=0$. 
With these definitions,
the scattering amplitude $\Mcal$, defined in Eq.~\eqref{Ttotdef}, satisfies the unitarity relation.
It is important to note that from the dressed propagator, defined in Eq.~\eqref{TRdef}, one can
infer a spectral function in the standard way via
\begin{equation}
\sigma_\res(s) = -\frac{1}{\pi}\Im G_\res(s)  \, ,
\label{spectraldef}
\end{equation}
which is automatically normalized 
\begin{equation}
\int_{s_{\thr}}^\infty \diff s \ \sigma_\res(s) = 1 \, .
\label{spectralnorm}
\end{equation}
This normalization condition is violated when the $s$-dependence of the real part of 
$\Sigma(s)$ that comes from the dispersion integral of Eq.~\eqref{Sigmadef} is
abandoned.

\begin{table*}[t]
   \centering\renewcommand{\arraystretch}{1.3}
   \caption{Parameters determined in the different analyses for the $\rho(770)$
   as well as the resulting values for the residues.
Note that the pole location is reproduced exactly by construction; cf.\  Eq.~\eqref{fixinggm2}.
   The uncertainties of the bare parameters reflect the impact of the uncertainties in the input parameters, for $s_B=1/f_R=\Lambda^2$, $\Lambda=2\GeV$ (upper) and $\Lambda=3\GeV$ (lower). For phase and modulus of the couplings, the first uncertainty refers to the impact of the uncertainties of the input parameters, the second one to the variation for $\Lambda\in[2,3]\GeV$ (in scenarios $(iii)$ and $(iv)$, $\tilde g_{\rho\pi\pi}$ is reproduced exactly, by construction).  
   Values marked with an asterisk are kept fixed in the fit.
   \label{tab:rho_params}}
   \begin{tabular}{ c c c c c  l l l l   }
   \toprule
  & $g$  & $m  \ \mbox{[GeV]}$ & $f_0$ [GeV$^{-2}$] & $f_1$ 
  & $|\tilde g_{\rho\pi\pi}|$ & $\mbox{arg}(\tilde g_{\rho\pi\pi}) \ [^\circ]$ &$|\tilde g_{\rho\bar KK}|$ &$\mbox{arg}(\tilde g_{\rho\bar KK}) \ [^\circ]$ \\ 
  \midrule
   \multirow{2}{*}{$(i)$} &   $6.61(2)$ & $0.84(5)$ & $0^*$ & $0^*$ &  \multirow{2}{*}{$5.95(6)(1)$} & \multirow{2}{*}{$-5.9(1.0)(0.7)$} & & \\ 
   &   $6.66(1)$ & $0.86(1)$ & $0^*$ & $0^*$ &  &  & & \\ 
   \hline
  \multirow{2}{*}{$(ii)$} &   $6.5(1)$ & $0.85(0)$ & $5(5)\times 10^{-6}$ & $0^*$ &  \multirow{2}{*}{$5.98(5)(1)$}  & \multirow{2}{*}{$-5.3(9)(1)$} & &\\ 
  &   $6.4(1)$ & $0.87(1)$ & $10(5)\times 10^{-6}$ & $0^*$ &   &  & &\\
  \hline
  \multirow{2}{*}{$(iii)$} &   $5.7(3)$ & $0.85(2)$ & $2.9(1.0)\times10^{-5}$ & $-1.6(4)$ &  \multirow{2}{*}{$6.01(3)$} & \multirow{2}{*}{$-5.3(1.0)$} & &\\
  &   $5.8(6)$ & $0.91(6)$ & $3.0(1.6)\times10^{-5}$ & $-1.0(4)$ &   &  & &\\
  \hline
   \multirow{2}{*}{$(iv)$} &   $5.9(5)$ & $0.84(4)$ & $2.6(1.1)\times10^{-5}$ & $-1.6(5)$ &  \multirow{2}{*}{$6.01(4)$}  & \multirow{2}{*}{$-5.3(9)$}  & \multirow{2}{*}{$3.3(3)(4)$} & \multirow{2}{*}{$-8.0(8)(8)$}\\
   &   $6.1(1.3)$ & $0.94(15)$ & $3.3(1.6)\times10^{-5}$ & $-1.1(4)$ &   &   &  & \\
  \bottomrule
   \end{tabular}
\end{table*}

\section{Generalization to higher partial waves}
\label{sec:higherpw}

To extend the parameterization outlined above to partial waves with $\ell>0$,
centrifugal barrier factors that grow as $q^\ell = (s/4-M^2)^{\ell/2}$ for small $s$ need to be included. However, as is
demonstrated, e.g., in Ref.~\cite{Weinberg:1995mt}, to be consistent with the positivity requirements of field theory, the physical propagator of a state is not allowed to drop faster than $1/s$ for large values of $s$.
Accordingly, Eq.~\eqref{TRdef} tells us that the self energy $\Sigma(s)$ is not allowed to grow faster than $s$ for all values of $\ell$. Thus, the energy dependence of the centrifugal barrier factors needs to be tamed.
Following Ref.~\cite{VonHippel:1972fg} we introduce the functions
\begin{equation}
\xi_\ell(s) = \sqrt{\frac{(s-4M^2)^{\, \ell}}{{2\ell +1}}} B_\ell\bigg(\frac{s-4M^2}{s_{\bg}-4M^2}\bigg) \, ,
\end{equation}
with the leading $B_\ell(x)$ given by 
\begin{align}  
   B_0=1\, , \qquad  
   B_1(x)=\sqrt{1/(1+x)} 
   \label{eq:bw-factor}\, .
\end{align}
Explicit forms for barrier factors with values of $\ell$ up to 4 are given, e.g., in Ref.~\cite{Chung:1995dx}. 
Here $s_\bg$ denotes some properly chosen scale with $s_\bg>4M^2$. 
The final results should not depend strongly on this parameter;
for definiteness we choose $s_\bg=1/f_R=\Lambda^2$, $\Lambda=2\GeV$, in the analyses below.
This is the scale that gives the best results for the naive resonance model
without a background interaction, see row $(i)$ of Table~\ref{tab:rho_params}, but we again include the variation to $\Lambda=3\GeV$ in the final uncertainty estimates.  
These regulator functions introduce unphysical singularities for space-like
values of $s$. However, since they are far away for the given choice of 
parameters and pushed to the unphysical sheet by construction in both the vertex functions and the self energies, they have no significant effect on the resonance
parameters and line shapes.

 In this work we restrict ourselves to 
systems of two spinless particles such that the total angular momentum is
captured in $\ell$. Then we can adapt the expressions
from above to the case $\ell\neq 0$ by employing
\begin{equation} 
\Sigma_\ell(s) = \frac{s-s_0}{\pi}\int_{s_\thr}^\infty \frac{\diff s'}{s'-s_0}\frac{\rho(s')\xi_\ell(s')^2|\gamma_\ell(s')|^2}{s'-s}  \, .
\label{Sigma_ell}
\end{equation}
The expression for $\Pi_\ell(s)$ follows from the one above by putting $\gamma_\ell(s)$ to 1 and choosing $s_0=4M^2$.
The vertex functions $\gamma_\ell(s)$ are still evaluated from Eq.~\eqref{gammaomnes}; however,
they need to be constructed from the phase of the adapted $\Mcal_\bg$, which now reads
\begin{equation}
\Mcal_\bg=\frac{\xi_\ell(s)^2f_0}{f(s)-f_0\Pi_\ell(s)} \, .
\end{equation}
For the resonance amplitude and the production amplitude we  find
\begin{align} 
  \Mcal_\res(s) &=  - g^2\gamma_\ell(s)^2\xi_\ell(s)^2 \, G_\res(s)  \notag\\
  &= -\frac{g^2\gamma_\ell(s)^2\xi_\ell(s)^2}{s-m^2+g^2\Sigma_\ell(s)} \, , \notag\\ 
\Acal_\res(s) &= - g \gamma_\ell(s)\xi_\ell(s)  \, G_\res(s) \, \alpha \, .
\end{align}
With this definition for the resonance propagator $G_\res(s)$, the spectral function introduced in Eq.~\eqref{spectraldef} remains normalized
according to Eq.~\eqref{spectralnorm} for all values of $\ell$. In contrast, using the expression for
$G_\res(s)$ provided in Ref.~\cite{Gounaris:1968mw}, the resulting spectral function is
not normalized due to the missing barrier factors $B_\ell(s)$, leading to
a resonance propagator that drops as $1/(s\log(s))$ for large values of $s$. One key advantage of our formalism is that the resulting spectral function is automatically normalized, which is not the case when improving Breit--Wigner-type parameterizations of the imaginary part of a resonance propagator via a dispersion integral~\cite{Lomon:2012pn,Moussallam:2013una,Zanke:2021wiq,Crivellin:2022gfu}. In this sense, we obtain a more direct implementation of the corresponding K\"all\'en--Lehmann spectral representation~\cite{Kallen:1952zz,Lehmann:1954xi} for a given resonance.

\section{Application to the $\rho(770)$}
\label{sec:rho770}

Before generalizing the formalism to coupled channels, we illustrate its application to the $\rho(770)$ and $f_0(500)$ resonances, respectively.
Pole parameters with very
high accuracy are available, e.g., from Refs.~\cite{Garcia-Martin:2011nna,Hoferichter:2023mgy}:
\begin{align}  
M_\rho &= 762.5(1.7)  \MeV \, , \notag \\
\Gamma_\rho &= 2\times 73.2(1.1) \MeV  \, , \notag \\ 
\tilde g_{\rho\pi\pi} &= 6.01(8)\exp{- i  \frac{\pi}{180}5.3(1.0)} \, . \label{rhoparameters}
\end{align}

\begin{figure*}[t!]
\centering
\includegraphics[width=0.49\textwidth]{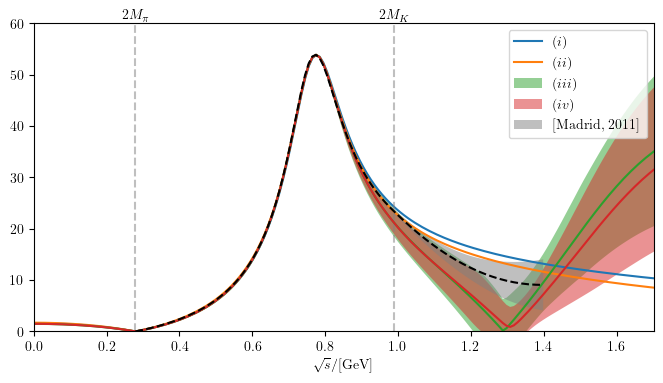}\hspace*{0.01\textwidth}
\includegraphics[width=0.49\textwidth]{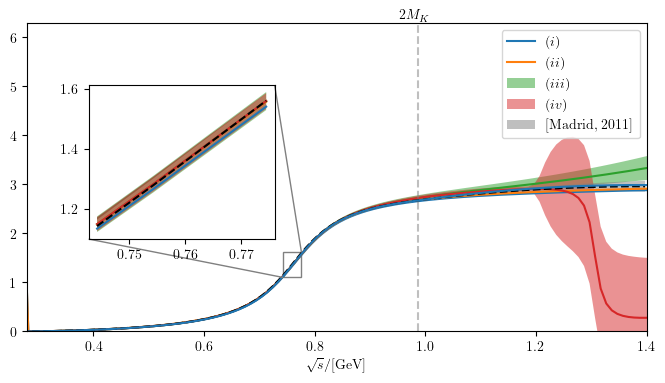}
\caption{The left (right) figure shows the absolute value (phase) of the $\pi\pi$ scattering amplitude, in both cases for the various analyses presented here:  
 $(i)$,  $(ii)$, $(iii)$, and $(iv)$ are shown as the blue, orange, green, and red line or band, respectively. 
In both figures we also show for comparison the results from Ref.~\cite{Garcia-Martin:2011iqs}. 
\label{rhodelandT}}
\end{figure*}

\begin{figure*}[t!]
\centering
\includegraphics[width=0.49\textwidth]{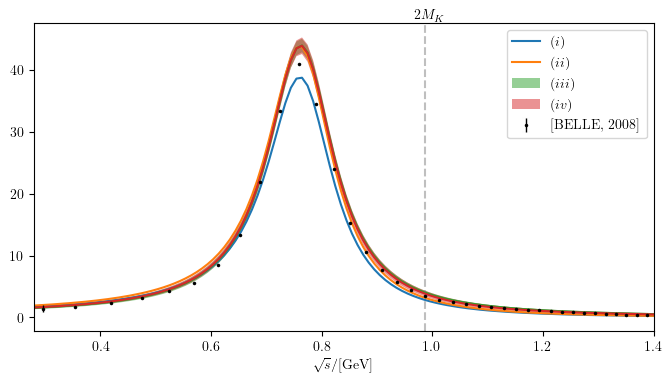}\hspace*{0.01\textwidth}
\includegraphics[width=0.49\textwidth]{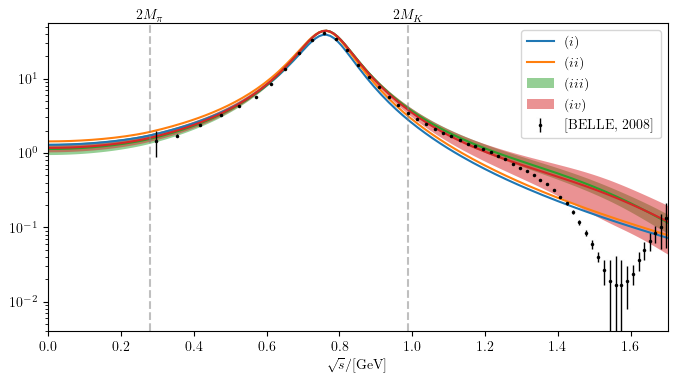}
\caption{The pion vector form factor compared to
data derived from $\tau^-\to\pi^-\pi^0\bar \nu_\tau$~\cite{Belle:2008xpe} on a linear scale (left) and on a logarithmic scale (right). Legend as in Fig.~\ref{rhodelandT}. \label{pipiPFF}}
\end{figure*}

The vector form factor is defined
via the current matrix element
\begin{equation}
\bra{\pi^+(p_1)\pi^-(p_2)}j_{\mu}^{(I=1)}\ket{0}=(p_1-p_2)_{\mu}F_\pi^V(s)\, ,
\end{equation} 
where $j_\mu^{(I=1)} = (\bar{u}\gamma_\mu u-\bar{d}\gamma_\mu d)/2$ and $s=(p_1+p_2)^2$. In the formalism introduced above it takes the form 
\begin{equation}
\label{FpiV}
    F_\pi^V(s)=B_{1}\bigg(\frac{s-4M_\pi^2}{s_B-4M_\pi^2}\bigg)\frac{-\alpha g \gamma_1(s)}{s-m^2+g^2\Sigma_1(s)}\, ,
\end{equation}
with the barrier factor $B_1(x)$ defined in Eq.~\eqref{eq:bw-factor}. 
This allows us to determine $\alpha$ in Eq.~\eqref{FFdef} via the coupling of the $\rho(770)$ to the photon 
\begin{equation}
\tilde g_{\rho \gamma}=5.01(7)\exp{-i\frac{\pi}{180}1(1)}\,
\end{equation}
as provided in Ref.~\cite{Hoferichter:2023mgy}. We emphasize that Eq.~\eqref{FpiV} does not yet define a suitable parameterization for precision studies of the pion vector form factor, for the following reasons: first, $F_\pi^V(s)$ is not normalized exactly to $F_\pi^V(0)=1$, since we only included the pole position and residues of the $\rho(770)$ as constraints, and this minimal parameterization violates the normalization by about $5\%$. Second, the barrier factor $B_1$ ensures a normalized spectral function, but introduces an unphysical LHC starting at $s=-(s_B-8M_\pi^2)$. Accordingly, the dispersion relation for $\Re F_\pi^V(s)$ in the physical region around the $\rho(770)$ is violated by $(2$--$3)\%$, a reasonably small effect given the scale $s_B\gtrsim 4\GeV^2$. Third, $F_\pi^V(s)$ behaves asymptotically as $1/s^{3/2}$, in contradiction to the expected $1/s$ scaling~\cite{Chernyak:1977as,Farrar:1979aw,Efremov:1979qk,Lepage:1979zb,Lepage:1980fj}. These shortcomings can be remedied by extending Eq.~\eqref{FpiV} appropriately, 
using the freedom in the choice of barrier factors and taking  into account polynomial terms in the unitarity relation for $F_\pi^V(s)$. Such generalizations will be studied in future work, while here we show the results for the minimal form~\eqref{FpiV}.

To demonstrate the effect of the background amplitude on the properties and
line shape of the $\rho(770)$ we applied three variants thereof: $(i)$ $\Mcal_\bg\equiv 0$,
$(ii)$ $k_{\rm max} = 0$, and $(iii)$ $k_{\rm max} = 1$, where the parameter $k_{\rm max} $,
introduced in Eq.~\eqref{fofsdef}, counts the number of terms in the expansion
in the conformal variable.

Since the parameters of the background amplitude ${\cal M}_{\rm B}$ enter the
expression for the resonance amplitude through an integral that can only be performed numerically, 
it is not possible to calculate the residue and its phase directly from the model parameters.
We therefore fit its parameters to the residue, while at all times demanding
\begin{align}
   \Im s_\res&= -g^2\Im\left(\Sigma_\ell^{(-)}(s_\res)\right) \, , \notag \\
   \Re s_\res&= m^2-g^2 \Re\left(\Sigma_\ell^{(-)}(s_\res)\right) \, ,
   \label{fixinggm2}
 \end{align}
where in case of the $\rho(770)$ discussed in this section we have $\ell=1$.
In this way the correct pole location is guaranteed.
The results of the three different
analyses are shown in Table~\ref{tab:rho_params}.
The uncertainties quoted in the table were determined via a bootstrap method,
where both residue and pole location were varied within their allowed uncertainties in the course of the analysis---always demanding that
there be no additional singularities  appearing in the amplitude. It is the
latter condition that leads to a slightly smaller uncertainty in the deduced residues than in the input residue. This limitation could be overcome by allowing for more parameters in the conformal expansion, to extend the region that can be scanned in the bootstrap procedure, but we restrict the analysis to the minimal case in which all parameters can be determined directly from the residues. 

\begin{table*}[t]
   \centering\renewcommand{\arraystretch}{1.3}
   \caption{
   Parameters determined in the different analyses for the $f_0(500)$
   as well as the resulting values for the residues. 
   Note that the pole location is reproduced exactly by construction; cf.\  Eq.~\eqref{fixinggm2}.
   The uncertainties of the bare parameters reflect the impact of the uncertainties in the input parameters, for $f_R=1/\Lambda^2$, $\Lambda=2\GeV$ (upper) and $\Lambda=3\GeV$ (lower). For phase and modulus of the couplings, the first uncertainty refers to the input parameters, the second one (where applicable) to the variation for $\Lambda\in[2,3]\GeV$.
   Values marked with an asterisk are kept fixed in the fit.  }
   \begin{tabular}{ c c c c c c  l l l l }
   \toprule
  & $g$  & $m$ & $f_0$ & $f_1 $ & $g_{\sigma\bar{K}K}$ & $|\tilde g_{\sigma\pi\pi}|$ & 
 $\mbox{arg}(\tilde g_{\sigma\pi\pi}) \ [^\circ]$ &$|\tilde g_{\sigma\bar KK}|$ &$\mbox{arg}(\tilde g_{\sigma\bar KK}) \ [^\circ]$  \\ 
  & $\mbox{[GeV]}$ & $\mbox{[GeV]}$ & & & $\mbox{[GeV]}$ & $\mbox{[GeV]}$ & 
  & $\mbox{[GeV]}$ &  \\ \midrule  
   $(i)$   &   $3.0(1)$ & $0.14(3)$ & $0^*$ & $0^*$ & $0^*$ &  $3.12(18)$  & $10(5)$ &&\\ 
 \hline
  \multirow{2}{*}{$(ii)$}  &   $5.3(5)$  & $0.89(6)$  & $-25.5(1.1)$ & $0^*$ & $0^*$ &  \multirow{2}{*}{$3.33(17)(8)$} &  \multirow{2}{*}{$-73.1(2.4)(0.5)$} &&\\ 
    &   $7.5(2.2)$  & $1.15(15)$  & $-27.5(1.3)$ & $0^*$ & $0^*$ &  &   &&\\ 
    \hline
  \multirow{2}{*}{$(iii)$} &   $4.7(3)$ & $0.82(3)$ & $-24.8(6)$ & $0.06(2)$ & $0^*$ &  \multirow{2}{*}{$3.61(11)$}  & \multirow{2}{*}{$-74.0(2.2)$} &&\\
  &   $5.3(4)$ & $0.82(5)$ & $-24.8(6)$ & $0.06(2)$ & $0^*$ &   &  &&\\
  \hline
  \multirow{2}{*}{$(iv)$}  &   $4.8(2)$ & $0.80(3)$ & $-24.9(4)$ & $0.06(2)$ & $2.3(2)$ & \multirow{2}{*}{$3.61(12)$} & \multirow{2}{*}{$-74.0(2.3)$} & \multirow{2}{*}{$2.0(1)$} & \multirow{2}{*}{$-23(1)(1)$}\\
  &   $5.4(6)$ & $0.85(8)$ & $-26.2(7)$ & $0.07(3)$ & $2.4(2)$ &  &  & & \\\bottomrule
   \end{tabular}
   \label{tab:sigma_params}
\end{table*}

The results show that in case of the $\rho(770)$ already the model without any background
gives a reasonable prescription of the residue. This should not come as a surprise, given that
the resulting amplitudes are very close to the Gounaris--Sakurai parameterization~\cite{Gounaris:1968mw}, with
the only difference that we employ the barrier function $B_\ell$, which, however, has a minor
impact in the resonance region.
As soon as we allow for a background amplitude,
both the absolute value and phase of the
residue can be exactly reproduced.
The resulting $\pi\pi$ $P$-wave scattering amplitude and the pion vector form factor are shown in Figs.~\ref{rhodelandT} and~\ref{pipiPFF}, respectively.

Two features of our analysis are worth noting.  First, the energy dependence of both the
scattering amplitude and the form factor are reproduced with rather high accuracy in all analyses, just using the correct $\rho$ pole parameters as well as the correct analytic structure of the amplitudes.  Second, as becomes evident in the right panel of Fig.~\ref{pipiPFF}, in the analysis $(iii)$, which reproduces the central values of the $\rho$ pole parameters exactly, the background amplitude introduces some resonance-like structure right in the mass range of the $\rho'$, which might indicate that the deviation of the $\rho$ pole parameters from the most naive implementation of the resonance physics realized in analysis $(i)$ is driven by the excited vector states.

\section{Application to the $f_0(500)$}
\label{sec:f0500}

\begin{figure*}[t!]
\centering
\includegraphics[width=0.49\textwidth]{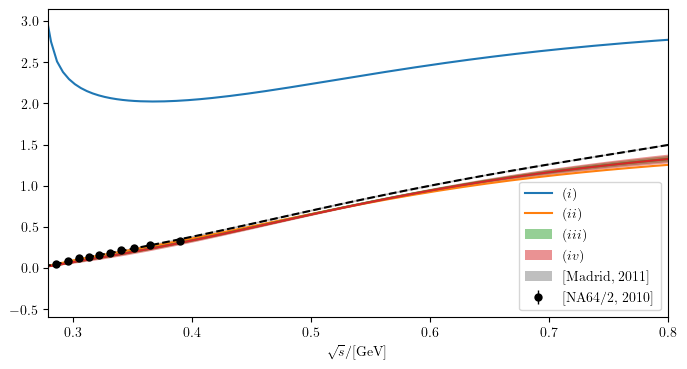}\hspace*{0.01\textwidth}
\includegraphics[width=0.49\textwidth]{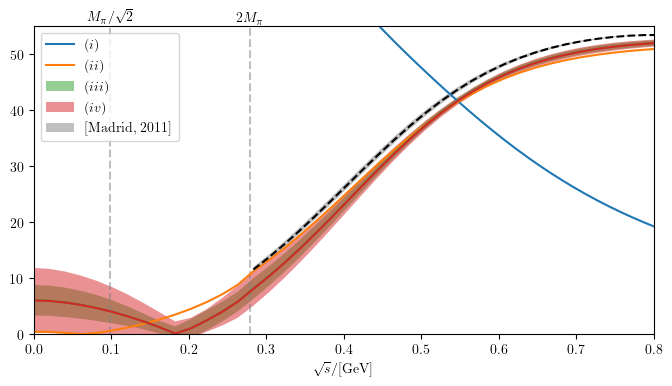}
\caption{Comparison of the phase shifts (left) and the absolute value of the scattering amplitude (right) 
that result for the scalar--isoscalar $\pi\pi$ channel, once the pole parameters are fixed via the  different variants
of the model:  
 $(i)$, $(ii)$, $(iii)$, and $(iv)$ are shown as the blue, orange, green, and red line or band, respectively.  
 The black dashed line shows the phase shift
 and the related absolute value of the scattering amplitude (between $\pi\pi$ and $\bar KK$ threshold) 
 from Ref.~\cite{Ananthanarayan:2000ht} for comparison. The dots show the phase shifts
 extracted from $K_{e4}$ decays~\cite{NA482:2010dug}. The first and second perpendicular lines show the locations
 of the Adler zero and the $\pi\pi$ threshold, respectively. 
\label{sigdelandT}}
\end{figure*}

\begin{figure}[t!]
\centering
\includegraphics[width=0.48\textwidth]{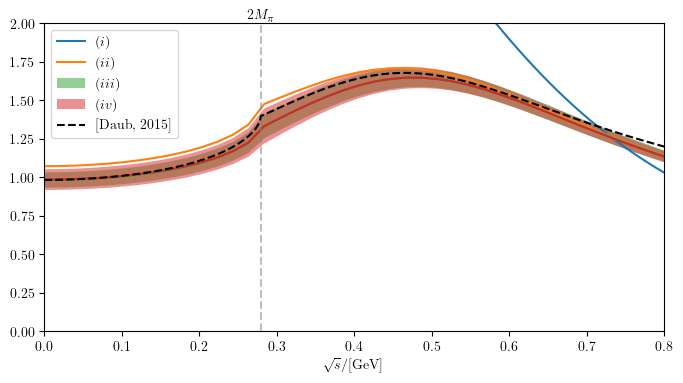}
\caption{Two-pion production amplitude that results for the different fits. The color code agrees with that of Fig.~\ref{sigdelandT},
only that now the dashed line is the non-strange scalar pion form factor of  Ref.~\cite{Daub:2015xja}, derived
from the phase shifts of Ref.~\cite{Ananthanarayan:2000ht}.}
\label{sigFFs}
\end{figure}

For the scalar--isoscalar channel, we use the pole parameters~\cite{Garcia-Martin:2011nna,Hoferichter:2023mgy}
\begin{align}  
M_\sigma &= 458(14)  \MeV \, , \notag \\
\Gamma_\sigma &=\ 2\times261(10) \MeV  \, , \notag \\ 
\tilde g_{\sigma\pi\pi} &= 3.61(13)\exp{-i \frac{\pi}{180}74(3)} \GeV\, . \label{sigmaparameters}
\end{align}
The scalar form factor is defined via the current matrix element
\begin{equation}
\bra{\pi\pi}j_{S}\ket{0}=M_{\pi}^2 F_\pi^S(s)\,,
\end{equation}
with the scalar current $j_{S}=\hat{m}(\bar uu+\bar dd)$.
In the formalism outlined above it takes the form
\begin{equation}
    F_\pi^S(s)=\frac{-\alpha g \gamma(s)}{s-m^2+g^2\Sigma(s)}\label{eq:explicitFS}\, .
\end{equation}
This allows us to determine $\alpha$ in Eq.~\eqref{eq:explicitFS} via the coupling of the $f_0(500)$ to a scalar source 
\begin{equation}
\tilde g_{\sigma S}=151(5)\exp{-i \frac{\pi}{180}25(2)}\MeV 
\end{equation}
as given in Ref.~\cite{Hoferichter:2023mgy}, see~\ref{app:conventions}.

As in the case of the $\rho(770)$, we perform three different analyses, with different levels of 
sophistication for the background amplitude.
As before, in all cases  the 
pole locations are reproduced exactly by 
employing the $S$-wave version of Eq.~\eqref{fixinggm2}. The results are reported in Table~\ref{tab:sigma_params}.
As one can see, in the absence of a background the residue of the $f_0(500)$ is not well described;
in particular, the phase of the residue is off completely. Also the resulting phase shifts and amplitudes
have little in common with our empirical knowledge of the scalar--isoscalar $\pi\pi$ amplitude, cf.\ the blue curves in Figs.~\ref{sigdelandT} and \ref{sigFFs}. In this case the resonance
amplitude acquires  an additional pole
right below threshold on the first sheet, in contradiction to the physical $\pi\pi$ scattering amplitude.
These observations reflect the fact that the features of the $f_0(500)$ cannot
be captured by a Breit--Wigner function, even
if an energy-dependent width is included.
The situation improves drastically when we allow for the simplest background amplitude, and especially as soon as the LHC is included in the parameterization the 
residue is reproduced exactly, in line with our modern understanding of the physics of the $f_0(500)$ resonance~\cite{Xiao:2000kx,Pelaez:2015qba}. 
With a non-vanishing background included in the analysis, fit $(ii)$, 
phase and absolute value of the residue are improved significantly. 
At the same time
the unphysical pole disappears and is replaced by a zero in the amplitude right below threshold, see the orange lines in 
Figs.~\ref{sigdelandT} and \ref{sigFFs}. When we include the $f_1$-term, 
the phase and absolute value of the residue are reproduced exactly. Also for this parameterization we
find a zero in the amplitude in
the same energy range. This is illustrated by the green curves in Figs.~\ref{sigdelandT} and \ref{sigFFs}.

The features described above demonstrate the intimate relation between the properties of the $f_0(500)$
and a non-trivial energy dependence of the $\pi\pi$ scattering amplitude in the threshold region.
In fact, since pions are the Goldstone bosons of the spontaneously broken chiral symmetry of QCD, there is necessarily a zero just below threshold in the $S$-wave isoscalar $\pi\pi$ scattering amplitude, the Adler zero~\cite{Adler:1964um,Adler:1965ga},
which at leading order (LO) in ChPT is located at $s_A=M_\pi^2/2$; as we show in~\ref{app:Adler}, this prediction is remarkably stable towards high-order corrections, with the one- and two-loop contributions reducing the LO value by $12\%$ and $3\%$, respectively.  
Since we derive the amplitudes from the pole parameters only, it should not come as a surprise
that the threshold physics driven by the chiral properties of QCD is not exactly reproduced,
however, it is a remarkable observation that reproducing the pole properties 
of the $f_0(500)$ with an amplitude consistent with unitarity and analyticity seems to
be possible only with amplitudes that 
feature a zero in the scattering amplitude
just below threshold, which finds a natural
explanation in the approximate chiral symmetry of QCD. The zeros found in the amplitudes are $s_A^{(ii)}=0.09\,\mpi^2$, $s_A^{(iii)}=2.1(4)\,\mpi^2$, and $s_A^{(iv)}=2.1(1.1)\,\mpi^2$.
While it has been known for a long time that unitarizing an amplitude that is in
line with chiral constraints leads to a pole of the $S$-matrix in close
proximity to that of the $f_0(500)$~\cite{Dobado:1996ps,GomezNicola:2007qj},
what we have demonstrated here for the first time is the opposite direction.

Finally, the black dashed lines in the two panels of Figs.~\ref{sigdelandT} and~\ref{sigFFs} show the correct phase shifts, absolute value of the scattering
amplitude, and production amplitude, respectively. They are based on the high-precision phase shifts of Ref.~\cite{Ananthanarayan:2000ht}.
The plots clearly illustrate that as soon as we include the background interaction, the qualitative features of
the $\pi\pi$ amplitude are reproduced reasonably well.

\section{Generalization to coupled channels and branching fractions}
\label{sec:branchings}

\subsection{Coupled channels}

For a meaningful discussion of branching fractions we need to extend the formalism introduced above to multiple channels, whose number shall be denoted by $n_c$. Since the goal of this study is to deduce line shapes from resonance properties and the residues factorize, it appears justified 
to introduce the background amplitude in diagonal form; however, the generalization
to a non-diagonal background is straightforward~\cite{Ropertz:2018stk}. Thus we 
write
\begin{equation}
\Mcal_\bg(s)_{ad} = \delta_{ad}\frac{\xi_{a}^2(s)f_{0\, a}}{f_a(s)-f_{0\, a}\Pi_a(s)} \, ,
\end{equation}
where $a,d\in \{1,\ldots,n_c\}$, $\Pi_a(s)$ denotes the non-interacting renormalized self energy 
of channel $a$, and $\xi_{a}(s)$ is the corresponding, channel-specific centrifugal barrier factor. In the multi-channel case the channel label not only specifies the
particle content of the given channel, but also
the angular momentum.  
Then the expression for the physical propagator 
reads
\begin{equation}
  G_\res(s)= \bigg(s-m^2+\sum_{a=1}^{n_c} g_a^2\Sigma_a(s)\bigg)^{-1} \, ,
  \label{TRdef_cc}
\end{equation}
and we obtain
\begin{align} 
\Mcal_R(s)_{ad} &= -  \xi_a(s)\gamma_a(s) g_a \, G_\res(s)  \, g_d  \gamma_d(s) \xi_d(s)   \, , \notag\\
\Acal_\res(s)_a &= - g_a \gamma_a(s)\xi_a(s) \, G_\res(s) \, \alpha \, , \label{Mdef_cc}
\end{align}
where $\gamma_a(s)$ is the vertex function that emerges from the background amplitude
$\Mcal_\bg(s)_{aa}$ according to Eqs.~\eqref{Tbdef} and \eqref{gammaomnes}.
If we allowed for off-diagonal terms in $\Mcal_\bg(s)_{ad}$, also the self energies would 
acquire off-diagonal terms~\cite{Ropertz:2018stk}.

For $n_c$ coupled channels one is faced with $2^{n_c}$ Riemann sheets. 
The resonance propagator $G_\res(s)$ on some arbitrary sheet
can be written as
\begin{equation}
  G_\res^{(\sh_1,\ldots,\sh_{n_c})}(s) = \bigg(s-m^2+\sum_{a=1}^{n_c} g_a^2\Sigma_a^{(\sh_a)}(s)\bigg)^{-1} \, ,
  \label{TRdef_cc_sheets}
\end{equation}
where the index $\sh_a\in \{+,-\}$ specifies on which sheet with respect to
channel $a$ the self-energy function needs to be evaluated.
In the single-channel analysis, the real and imaginary parts of the pole location allowed us to determine
$m$ and $g^2$ for any given background; see Eq.~\eqref{fixinggm2}. The situation is a little more
complicated now, since various channels and the corresponding couplings appear in the denominator of 
the resonance propagator $G_\res(s)$ defined in Eq.~\eqref{TRdef_cc}. In practice,
the procedure to fix the proper pole location(s)
 depends on what information is available for the resonance under study. For example, if pole
locations on various sheets are known, one may straightforwardly generalize Eq.~\eqref{fixinggm2} to fix $g_a$ as well as $m$. If, however, only one pole is known, then Eq.~\eqref{fixinggm2} only determines one of the $g_a$ couplings and the others may be employed to fix the pertinent residues. 
In the examples discussed in this paper, the two-pion channels are by far dominating, and we therefore use Eq.~\eqref{fixinggm2} as given to fix $m$ and $g_{\pi\pi}$. 
The additional inclusion of the couplings to relatively
unimportant inelastic channels, like $\gamma\gamma$ for the
$f_0(500)$, only changes the pole location within uncertainties. Therefore, we will use for those the approximation 
\begin{equation}
   g^2_{a}=|\tilde g_{a}|^2/|Z| 
\end{equation}
to find the corresponding couplings. For more strongly coupled channels, such as the $\bar K K$ channel in the scalar--isoscalar 
$\pi\pi$ system, we implement the bare coupling as an additional free parameter in the fit to reproduce the residues.

\subsection{Branching fractions}
\label{sec:branching_fractions}

In this subsection we compare different possible definitions for
branching fractions and propose a new one based on the formalism
discussed in the previous sections.

In Eq.~\eqref{eqn:MGamdef} the total width of a resonance was fixed from the pole location $s_\res$
as $\Gamma_\res = -\Im s_\res /M_\res$.
Matching that to Eq.~\eqref{TRdef_cc_sheets}
reveals a natural definition of partial widths, namely
\begin{equation}
 \Gamma_a^{\res \rm (p)} = -g^2_a\Im(\Sigma_a^{(\sh_a)}(s_\res))/M_\res \, ,
  \label{Gamtotdef}
\end{equation}
where the index $(\sh_a)$ is fixed by the sheet on which the pole at $s_\res$ 
is located. Clearly such a definition is sensible only if it is just a single pole that dominates the physics, however, this holds true both for the $f_0(500)$ and the $\rho(770)$.
In the absence of background interactions for a single channel, this definition agrees with the
one of Ref.~\cite{Willenbrock:2022smq}.
For the evaluation of the branching ratio from the pole location, one then obtains
\begin{equation}
\BR_a^{\rm (p)} = \Gamma_a^{\res \rm (p)}/\Gamma_\res   \, .
\label{Brpoledef}
\end{equation}
However, it was shown in Ref.~\cite{Wang:2022vga} 
using the example of $f_0(980)$ and $a_0(980)$
(and further discussed for the former resonance in Ref.~\cite{Burkert:2022bqo}) that this definition runs into problems 
when the most relevant pole sits on a sheet other than the one adjoined to the physical sheet 
above all thresholds. Taking the $f_0(980)$ as an example, where the pole typically
sits above the $\bar KK$ threshold, but on the physical sheet with respect to the $\bar KK$ channel, the problem  is that the different contributions to the imaginary part of the pole location in 
Eq.~\eqref{TRdef_cc} no longer add up, since $\Im(\Sigma_{f_0(980)\bar KK}^{(+)}(s_{f_0(980)}))>0$
but $\Im(\Sigma_{f_0(980)\pi\pi}^{(-)}(s_{f_0(980)}))<0$. The authors argue that in this case $\Gamma_\res$ is 
not a proper measure of the total width. Adapting their insights to our parameterization, we define
a modified expression from the one given by the pole location:
\begin{align} 
 \Gamma_a^{\res \rm (p,m)} &= g^2_a\left|\Im(\Sigma_a^{(\sh_a)}(s_\res))/M_\res\right| \, , \notag \\ 
 \Gamma_{\rm tot}^\res &= \sum_a  \Gamma_a^{\res \rm (p,m)} \, ,  \notag \\ 
 \BR_a^{\rm (p,m)} &= \Gamma_a^{\res \rm (p,m)}/\Gamma_{\rm tot}^\res  \, . \label{Gamtotdefmod} 
\end{align}
It should be stressed that the mentioned sign problem highlighted for the example of the $f_0(980)$
is considerably more general: it occurs as soon as channels are included in the analysis 
that are closed at the resonance location, as is the case for the decay of both $f_0(500)$
and $\rho(770)$ to two kaons.
 Thus, we also include $\BR_a^{\rm (p,m)}$
in our study.

In some cases, such as the two-photon coupling of the $f_0(500)$, a narrow-width formula has been used in the literature to turn the residue into a decay rate~\cite{Morgan:1987gv,Morgan:1990kw,Hoferichter:2011wk,Moussallam:2011zg}.
Thus, in this prescription one has
\begin{equation}
\Gamma_a^{\res \rm (nw)}=\frac{|\tilde g_a|^2}{M_R}\rho_a(M_R^2)\xi^2_a(M_R^2) \, ,
\label{narrowgammadef}
\end{equation} 
where the superscript (nw) refers to the narrow-width limit.
To see how well this prescription works in practice, we define
\begin{equation}
\BR_a^{\rm (nw)} = \Gamma_a^{\res \rm (nw)}/\Gamma_\res  \, .
\label{Brnwdef}
\end{equation}
Contrary to the branching ratios defined in Eq.~\eqref{Brpoledef}, those of Eq.~\eqref{Brnwdef} not necessarily add to $1$.
Only for narrow states above threshold, where $M_\res^2-4M_a^2\gg M_\res\Gamma_\res$ holds, one finds
\begin{equation}
g^2_a\Im(\Sigma_a^{(-)}(s_\res))\approx -|\tilde g_a|^2 \rho_a(M_\res^2) \xi^2_a(M_R^2)\, ,
\label{eq:recovernw}
\end{equation}
and Eq.~\eqref{narrowgammadef} is recovered naturally. 
However, for $M_\res^2 - 4M_a^2<0$ the phase-space factor $\rho_a(M_R^2)$ vanishes and
thus for that case the narrow-width formula does not provide a meaningful answer. 

The definition we propose to use for the evaluation of branching fractions
is a lot closer to what is measured in
experiment for a single, isolated resonance. 
What is done there can be summarized as~\cite{Burkert:2022bqo}
\begin{equation}
\BR_a^{\rm exp} = N_a/N_{\rm tot} \, , \qquad N_{\rm tot} =\sum_{a=1}^{n_c} N_a \, , 
\label{eq:expbranchings}
\end{equation}
where $N_a$ is the number of events measured for the decay of the resonance R
into channel $a$ in some production reaction
(assuming that the resonance leaves a
sufficient imprint in the channel).
The specifics of the production reaction cancel in the ratio and thus $\BR_a^{\rm exp}$  measures a resonance
property.
Given that the count rates in a channel $a$ from some resonance $\res$ are calculable from $G_\res$, evaluated on the physical sheet
such that $\sh_a=+$ for all $a$ (the superindex introduced in Eq.~\eqref{TRdef_cc_sheets} is dropped here to ease notation),
we can write
\begin{equation}
  \BR_a^{\rm (cr)} =  \int_{s_{\thr}^a}^\infty \frac{\diff s}{\pi} \left|G_\res(s) g_a
  \gamma_a(s)\xi_{a}(s)\right|^2 \rho_a(s)\, ,
  \label{eq:brcrdef}
\end{equation}
where the label (cr) shows the relation to the count rates. Since the formalism described above automatically generates a spectral
function that is normalized, the sum over the $\BR_a^{\rm (cr)}$ is one. Moreover, all the self energies need 
to be evaluated on the physical, the $(+)$, sheet and accordingly no sign problem can appear,
regardless of where the pole is located.

The construction Eq.~\eqref{eq:brcrdef} encodes the properties of a single resonance.
As long as pole locations and residues are known with sufficient accuracy for each individual state, it should also be applicable for partial waves with various overlapping resonances, although in this case the method sketched in Eq.~\eqref{eq:expbranchings} can no longer be applied to experimental data straightforwardly.  
Concrete tests hereof, including the sensitivity to the parameterization of background amplitude and barrier factors as studied here for the $\rho(770)$ and $f_0(500)$,  are left for future work.

\begin{table*}[t]
   \centering\renewcommand{\arraystretch}{1.3}
    \caption{Comparison of the branching ratios calculated using the different prescriptions introduced in Sec.~\ref{sec:branching_fractions}. The dagger indicates that for those branching fractions the uncertainties could not be evaluated, for the reasons detailed in the main text. Whenever two uncertainties are 
    provided, the first one refers to that in the input quantities, the second one to the variation for $\Lambda\in[2,3]\GeV$; see Tables~\ref{tab:rho_params} and \ref{tab:sigma_params}. 
    \label{Brresults}}
   \begin{tabular}{ c c c c c c  c  c }
   \toprule
   & & narrow width   & \multicolumn{2}{c}{pole location} & \multicolumn{2}{c}{Ref.~\cite{Burkert:2022bqo}} & this work\\ 
Resonance & channel ($a$)  & $\BR_a^{\rm (nw)}$  & $\BR_a^{\rm (p)}$  & $\BR_a^{\rm (p,m)}$  & $\BR_a^{\rm (B)}$ & $\BR_a^{\rm (B,n)}$  & $\BR_a^{\rm (cr)}$\\ 
\midrule
$f_0(500)$ & $\pi\pi$  & $0.8(1)$ & $1.03(1)(0)$ & $0.97(0)(0)$ & $0.52(8)$ & $0.94(5)$ & $0.970(5)(12)$ \\
 & $\gamma\gamma \ \times 10^{6}$   
 & $3.0(7) $ 
 & $5.0(1.6)(0.3) $ 
 & $4.8(1.5)(0.3) $ 
 & $1.9(6) $ 
 & $3.5(9) $
 & $1.4(4)(3) $ \\
 & $\bar KK$   & $0$   & $-0.03(1)(0)$ & $0.03(1)(0)$ & $0.03(2)$ & $0.06(4)$ & $0.030(5)(12)$ \\
 \midrule
 & sum   & $0.8$ & $1.0$ & $1.0$ & $0.74$ & $1.0$ & $1.0$ \\
\midrule
\midrule
$\rho(770)$ & $\pi\pi$  &  $1.007(14)$  &  $1.04(1)(4)$  & $0.96(1)(3)$   &   $1.222(5)$   & $0.967(3)$  &   $0.95(4)(3)$   \\   
   &   
 $\pi\gamma \ \times 10^{4}$ & $5.1(1.1)$ & $3(1)(6)$ & $3(1)(3)$ & $5.4^\dagger$ & $4.3^\dagger$ & $12(1)(4)$ \   \\   
   & $\bar KK$   & $0$   & $-0.05(1)(3)$   & $0.04(1)(3)$  &   $0.0419(5)$   &   $0.0331(3)$   & $0.05(4)(3)$   \\   
   \midrule
   & sum & $1.0$   &   $1$   &  $1$   &   $1.26$   &   $1$   &   $1$   \\   
   \bottomrule   
   \end{tabular}
\end{table*}

As a final definition, we compare our results to  
Eq.~(28) in combination with Eq.~(19) of Ref.~\cite{Burkert:2022bqo}:
\begin{equation}
 \BR_a^{\rm (B)} = \int_{s_{\thr}^a}^\infty \frac{\diff s}{\pi}
 \frac{f|\tilde{g}_a|^2\rho_a(s)\xi_{a}(s)^2}{\left|s-\hat{m}^2+i f\sum_b |\tilde{g}_b|^2 \rho_b(s)\xi_{b}(s)^2\right|^2} \, ,
 \label{BrBeadef}
\end{equation}
where as before the $\tilde{g}_a$ denote
the effective couplings derived from the
residues. However, as discussed in Sec.~\ref{sec:Swaves}, without additional background contributions it is not possible to simultaneously obtain both the correct pole location and residue. Because of this, the authors of Ref.~\cite{Burkert:2022bqo}
introduced a fudge factor, $f$, adjusted along with $\hat{m}^2$ in such a way that the pole location is correct.
Thus, 
Eq.~\eqref{BrBeadef}
 is close to what one would obtain in our formalism for a vanishing background, only that the 
dispersive pieces of the self energies are dropped. As discussed above, 
Eq.~\eqref{BrBeadef}
does in general not lead to the correct residues, however, 
for the $\rho(770)$, this is not necessarily a big effect. Moreover,
Eq.~\eqref{BrBeadef} relates to a spectral function that  is not normalized, resulting in branching fractions that do not sum to $1$.
Thus, to allow for a better comparison, we also introduce a normalized branching ratio based on Eq.~\eqref{BrBeadef},
namely
\begin{equation}
 \BR_a^{\rm (B,n)} = \frac{1}{N}\BR_a^{\rm (B)}\, , \qquad  N=\sum_{a=1}^{n_c}\BR_a^{\rm (B)} \, .
 \label{BrBeadef_norm}
\end{equation}

\subsection{Results for the $\rho(770)$ and $f_0(500)$}

Using the examples of the $\rho(770)$ and $f_0(500)$, we now compare the 
results of the various prescriptions to calculate branching fractions, including the dominant $\pi\pi$ decay as in Secs.~\ref{sec:rho770} and~\ref{sec:f0500}, but including as well  the 
$\gamma\gamma$ and $\bar KK$ channels for the $f_0(500)$~\cite{Hoferichter:2011wk,Danilkin:2020pak} 
\begin{align}
\tilde g_{f_0(500)\gamma\gamma}&= 6.3(7)\exp{-i\frac{\pi}{180}115} \MeV \, , \notag \\
 \tilde g_{f_0(500)\bar K K}&= 2.1(4)\exp{-i\frac{\pi}{180} 57.9} \GeV\, ,
\end{align}
and the $\pi\gamma$ and $\bar K K$ ones for the $\rho(770)$. The former has the residue~\cite{Hoferichter:2017ftn} 
\begin{equation}
\tilde g_{\rho\pi\gamma}=\sqrt{8\pi\alpha_{\text{em}}}\ 0.79(8)  \GeV^{-1} \, ,
\end{equation}
where $\alpha_{\text{em}}$ is the fine-structure constant,
with a phase consistent with zero.
Compared to the coupling of the pions to the $\rho(770)$, the dimension is different due to an additional momentum dependence in the vertex; see~\ref{app:radiative} for details.

The residue for the coupling of the $\rho(770)$ to $\bar KK$ is estimated using an SU(3) symmetric vector-meson-dominance Lagrangian~\cite{Meissner:1987ge,Klingl:1996by} and by comparing the vector--isovector $\pi\pi$ and $\bar{K}K$ form factors (cf.\ also Ref.~\cite{Stamen:2022uqh}). With that we can approximate the bare coupling of the $\rho$ to the kaons as $g_{\rho\bar{K}K}^2=g_{\rho\pi\pi}^2/2$ to obtain a prediction for the value of the residue.

The branching ratios calculated for these systems with the different methods are shown in Table~\ref{Brresults}. 
As expected from Eq.~\eqref{eq:recovernw}, for those cases in which the inelastic threshold is well below the resonance mass,
as is the case for the $\gamma\gamma$ decay of the $f_0(500)$ and the $\pi\gamma$ decay of the
$\rho(770)$,
the narrow-width formula, $\BR_a^{\rm (nw)}$, gives results (almost) consistent with the ones derived from the 
pole location, $\BR_a^{\rm (p)}$. However, 
some deviations are observed in comparison to $\BR_a^{\rm (cr)}$, which reflects the impact of the line shape on the branching fractions---note that $s_\thr^{\gamma \gamma}=0$
and $s_\thr^{\pi \gamma}=\mpi^2$,
so that the line shape is probed
over a large range when the integral
in Eq.~\eqref{eq:brcrdef} is evaluated.

The effective prescription from Ref.~\cite{Burkert:2022bqo},
$\BR_a^{\rm (B)}$,
suffers from the wrong normalization of Eq.~\eqref{BrBeadef}. Therefore, already the 
$\pi\pi$ branching ratio deviates significantly from the other cases. If one corrects for this, the agreement 
with $\BR_a^{\rm (cr)}$ improves, as shown in column $\BR_a^{\rm (B,n)}$.
However, employing $\BR_a^{\rm (B,n)}$ to calculate the two-photon width of the $f_0(500)$ gives
a result that is two standard deviations
larger than the reference value provided in the column
marked as $\BR_a^{\rm (cr)}$. This large discrepancy follows from 
the 
increased sensitivity to the line shape of the $f_0(500)$ at small values of $s$, which in this parameterization becomes similar to the blue solid
line in Fig.~\ref{sigFFs}. 

We were not able to determine the uncertainties for 
the $\pi\gamma$ branching fraction of the $\rho(770)$ 
for $\BR_a^{\rm (B)}$ and $\BR_a^{\rm (B,n)}$. The 
reason is that the integrand in Eq.~\eqref{BrBeadef} develops a pole below the two-pion threshold, since the analytic continuation
of the $\bar KK$ momentum becomes sizable here and contributes
negatively. The same problem does not occur for 
Eq.~\eqref{eq:brcrdef}, since here the analytic continuation
of the momentum is tamed by the dispersion integral. Furthermore, in the case of $s_B = (3\GeV)^2$, we observe that the imaginary part of the $\pi\gamma$ self energy on the second sheet at the $\rho(770)$ pole location changes sign compared to the central solution at $\Lambda=2\GeV$. Such zeros on the second sheet also occur for other channels, but the $\pi\gamma$ case is the only one for which we find a strong sensitivity of its position to the regulator scale. In contrast, the behavior on the real axis appears to be more stable, suggesting that indeed $\BR^{\rm (cr)}_a$ defines a better prescription for a branching fraction than $\BR^{\rm (p)}_a$.

When the threshold for the inelastic channel lies above the resonance
location, as for the $\bar KK$ decay of the $f_0(500)$ and $\rho(770)$,
the various expressions naturally give very different results,
and $\BR_a^{\rm (nw)}$ even becomes zero. However, also the prescription via the pole location that appears improved at first glance, $\BR_a^{\rm (p)}$, gives a negative value for both resonances, and thus does not produce a meaningful branching fraction in this case either.
 All other prescriptions give consistent results
within uncertainties. The uncertainty of the branching
fraction $\BR_{\rho\to\bar KK}^{\rm (cr)}$ is significantly
larger than all others; this reflects the fact that the
$\rho(770)$ line shape is badly determined for energies beyond $1\GeV$,
where it is probed for this channel. 

\section{Summary  and outlook}
\label{sec:summary}

In this paper we introduced a formalism consistent with the fundamental principles of analyticity, unitarity, and positivity of the spectral function that allows one to derive line shapes of a resonance 
solely from its pole parameters. The resulting
spectral function is automatically normalized, allowing for an unambiguous definition of branching ratios via proper integrals over
the given line shape. As test cases, we discussed the $\rho(770)$ and $f_0(500)$ resonances, whose pole parameters are known to high precision from dispersive analyses of $\pi\pi$ scattering. In particular, their study allowed us to assess which degrees of freedom are required to capture all relevant features of the respective resonance. 

For the $\rho(770)$ we found that a simple Dyson resummation of the self energy, essentially corresponding to a Gounaris--Sakurai parameterization, gives reasonable agreement with phenomenology, but is not sufficient to match the available precision, mainly because the residue is already determined by the pole position. Accordingly, we improved on the construction by including a background term in the two-potential formalism, which provides the required freedom to adjust the residue as well. We observed that the corresponding corrections seem to be concentrated in the energy range in which the excited $\rho'$ and $\rho''$ resonances appear, and could thus be interpreted as a hint where $4\pi$ effects become relevant~\cite{Hanhart:2012wi,Chanturia:2022rcz}.  A possible future application concerns the 
$2\pi$ contribution to hadronic vacuum polarization~\cite{Colangelo:2018mtw,Colangelo:2020lcg,Colangelo:2022prz}, given that the tensions among different data sets, most prominently BaBar~\cite{Lees:2012cj}, KLOE~\cite{Anastasi:2017eio}, and CMD-3~\cite{CMD-3:2023alj}, indeed appear to point to the study of inelastic effects as an important means to better understand the discrepancies~\cite{Colangelo:2023rqr}.  

For the $f_0(500)$, we found that, as expected, it is critical to account for the LHCs, which we implemented including the correct threshold behavior $\propto (-s)^{3/2}$. Moreover, we found that demanding the precise resonance pole position and residue automatically implies a subthreshold zero in the $\pi\pi$ scattering amplitude, which can be naturally identified with the Adler zero. In fact, while it is well known that unitarizing ChPT amplitudes with the Adler zero generates a pole close to the $f_0(500)$ found from Roy equations, our study shows that also the opposite is true. 
As a by-product, we evaluated the chiral corrections to the position of the Adler zero, finding a $15\%$ reduction compared to its LO value. 

Finally, the new way to evaluate spectral functions also allowed
us to introduce a new expression to calculate
branching fractions via integrals over resonance line shapes
that by construction contain information on the correct
pole location and residues. 
We compared our prescription to alternatives proposed in the literature. In some cases significant differences were observed and the origin of those was identified, e.g., related to (lack of) normalization of the spectral function and sensitivity to the line shape far away from the resonance. While the cases we studied are still dominated by the $\pi\pi$ channel, a major advantage of our proposed formalism is that it applies to situations in which different channels can compete, leading to a more complicated analytic structure. This includes the $f_0(980)$, with its strong interplay of $\pi\pi$ and $\bar K K$ $S$-waves, as well as the $a_0(980)$, in which case $\pi\eta$ and $\bar K K$ have comparable branching fractions. We leave the study of such systems to future work.

\begin{acknowledgements}
We thank Jos\'e~R.~Pel\'aez for useful discussions and
 the INT Seattle for its hospitality during the workshop
 {\it Accessing and Understanding the QCD Spectra} in March 2023,
 where parts of this work were completed, supported 
 by the INT's U.S.\ Department of Energy grant No.\ DE-FG02-00ER41132. We further thank 
 Eberhard Klempt, Mikhail Mikhasenko, and Jos\'e~A.~Oller for comments on the manuscript. 
This work is supported in part 
 by the MKW NRW under the funding code NW21-024-A;
 by the Konrad-Adenauer-Stiftung e.V.\ with funds from the BMBF; 
 by the Volkswagenstiftung (Grant No.~93562);
 by the Chinese Academy of Sciences under Grants No.~YSBR-101 and No.~XDB34030000;
 by the NSFC under Grants No.\ 12125507, No.\ 12361141819, and No.\ 12047503;
 and by the SNSF (Project No.~PCEFP2\_181117).
\end{acknowledgements}

\appendix

\section{Conventions}
\label{app:conventions}

An integral part of the work presented in this paper concerns the adjustment of parameters to reproduce the correct resonance parameters. Especially the residues are often subject to different conventions and definitions. Here, we collect the various conventions for amplitudes and form factors.  

The isoscalar--scalar amplitude on the second sheet close to the pole can be written in the form
\begin{equation}
   t^0_0(s)^\text{II} = \frac{g^2 Z \gamma^\text{II}(s_p)^2}{s_p-s} = \frac{1}{16\pi}\frac{{\tilde g_{\sigma\pi\pi}}^2}{s_p-s}\, ,
\end{equation}
with the renormalization factor $Z$. The form factor can be written in a similar form, 
\begin{equation}
   F_\pi^S(s)^\text{II} = \frac{g Z \alpha\gamma^\text{II}(s_p)}{s_p-s}
   = \frac{{\tilde g_{\sigma\pi\pi}} \sqrt{Z} \alpha}{s_p-s}\, .
\end{equation}
We can use this form to write an effective matching condition between the coupling of a scalar source to the $f_0(500)$ and our source coupling $\alpha$. Note that the change from the particle to the isospin basis causes an additional factor in our description, leading to
\begin{align}
   \bra{0}j_S\ket{f_{0}(500)}&=\bra{0}j_S\ket{I=0}=M_{\pi}^2\tilde g_{\sigma S}\notag\\
   &=M_{\pi}^2\sqrt{\frac{3}{2}}\frac{\text{Res}\, F^\text{II}_S}
   {\tilde g_{\sigma\pi\pi}} \notag \\
   &=M_{\pi}^2 \sqrt{\frac{3}{2}}\alpha\sqrt{Z}\, .
\end{align} 
The isovector--vector amplitude on the second sheet close to the pole can be written as
\begin{align}
   t^1_1(s)^\text{II}&= \xi^2(s_p)\frac{g^2 Z \gamma^\text{II}(s_p)^2}{s_p-s}\notag\\
   &=\frac{(s_p-4M_{\pi}^2)}{3}\bigg[B_{1}\left(\frac{s_p-4M_{\pi}^2}{s_B-4M_{\pi}^2}\right)\bigg]^2\frac{g^2 Z \gamma^\text{II}(s_p)^2}{s_p-s}\notag\\
   &= \frac{(s_p-4M_{\pi}^2)}{48\pi}\frac{{\tilde g_{\rho\pi\pi}}^2}{s_p-s}\, .
\end{align}
It should be noted that in this case the function $H(s)$ as defined in Eq.~\eqref{eqn:resdef} absorbs only the squared momentum and the normalization, but not the taming factor. Accordingly, the vector form factor can be written as
\begin{align}
   F_\pi^V(s)^\text{II}&= B_{1}\left(\frac{s_p-4M_\pi^2}{s_B-4M_\pi^2}\right)\frac{g Z \alpha\gamma^\text{II}(s_p)}{s_p-s}\notag\\
   &= \frac{\alpha \sqrt{Z} {\tilde g_{\rho\pi\pi}}}{s_p-s}\, ,
\end{align}
leading to
\begin{equation}
   \frac{\tilde g_{\rho\pi\pi}}{\tilde g_{\gamma\rho}}s_p=\alpha\sqrt{Z} {\tilde g_{\rho\pi\pi}}\, .
\end{equation}

\section{Electromagnetic channels and couplings}
\label{app:radiative}

In this work two channels were studied that are different from the massive pseudoscalar ones, $\pi\pi$ and $\bar K K$, i.e., the radiative decays into $\gamma\gamma$ and $\pi\gamma$ for the $f_0(500)$ and $\rho(770)$, respectively.  The couplings of these channels are suppressed by orders of $\alpha_{\text{em}}$ and the self-energy contributions have a different structure. Here, the definitions of the respective couplings  as well as their self-energy contribution are collected.

The width of the decay $f_0\to\gamma\gamma$ in the narrow-width limit is defined as
\begin{equation}
    \Gamma_{f_0\gamma\gamma}=\frac{e^4|\hat g_{f_0\gamma\gamma}|^2}{16\pi M_{f_0}}
    =\frac{|\tilde g_{f_0\gamma\gamma}|^2}{M_{f_0}}\Im\Pi_{\gamma\gamma}(M_{f_0}^2)\, .
\end{equation}
Therefore, the imaginary part on the real axis of the self-energy contribution is given as
\begin{equation}
    \Im\Pi_{\gamma\gamma}(s)=\frac{1}{16\pi}\, ,
\end{equation}
which can be used in a once-subtracted dispersion integral to determine explicitly 
\begin{equation}
    \Pi_{\gamma\gamma}(s)=\frac{1}{16\pi^2}\log\bigg(\frac{-s_{\gamma\gamma}}{s}\bigg)\, ,
\end{equation}
where $s_{\gamma\gamma}>0$ is the subtraction point.

The decay $\rho\to\pi^0\gamma$ needs to be treated more carefully, since its amplitude reads
\begin{equation}
\mathcal{M}_{\rho\pi\gamma}=e{\hat g}_{\rho\pi\gamma}\epsilon_{\mu\nu\alpha\beta}\epsilon_{\rho}^{\mu}\epsilon_{\gamma}^{\nu}p_{\pi}^{\alpha}p_{\gamma}^{\beta}\, .
\end{equation}
The width of the decay in the narrow-width limit is defined as
\begin{align}
   \Gamma_{\rho\pi\gamma}&=\frac{e^2|\hat g_{\rho\pi\gamma}|^2}{96\pi M_{\rho}^3}(M_{\rho}^2-M_{\pi}^2)^3\notag\\
   &= \frac{|\tilde g_{\rho\pi\gamma}|^2}{M_{\rho}}\Im\Pi_{\pi\gamma}(M_{\rho}^2)\, ,
\end{align}
and therefore
\begin{equation}
   \Im \Pi_{\pi\gamma}(s)=\frac{1}{16\pi}\frac{s^2}{12}\left(1-\frac{M_{\pi}^2}{s}\right)^3
\end{equation}
is used to define the self-energy contribution at the pole by calculating the respective  dispersion integral, once-subtracted at $s_0=0$. To tame the energy dependence we use the function $[B_1(x)]^4$, with $B_1(x)$ as defined in Eq.~\eqref{eq:bw-factor}.

\section{Two-body left-hand cuts}\label{app:LHC}

LHCs in partial-wave amplitudes are due to singularities in crossed ($t$-, $u$-) channels.  As they appear as a result of partial-wave projection, the associated integration over the scattering angle in general weakens the singularity: crossed-channel \textit{poles}, e.g., turn into left-hand \textit{cuts}.  In the context of crossing-symmetric $\pi\pi$ scattering, the leading LHC is again due to two-pion intermediate states, which lead to a branch point at $s=0$.  We discuss the degree of the corresponding singularity with the help of the one-loop function (subtracted at $s=0$)
\begin{align}
    \bar J(s) &= 2 - 2\sigma^2 L_\sigma \,, \quad L_\sigma = \frac{1}{2\sigma}\log\frac{\sigma+1}{\sigma-1} \,, \notag\\
    \sigma &= \sqrt{1-\frac{4M^2}{s}} = 16\pi\rho(s) \,,
\end{align}
where we have absorbed an overall factor in the definition of $\bar J$ for simplicity.
Its right-hand cut is of the well-known square-root type,
\begin{equation}
\Im \bar J(s) = \pi\,\sigma\,\theta\big(s-4M^2\big) \,.
\end{equation}
We now define $s$-channel partial-wave projections of $\bar J(t)$ according to
\begin{align}
    \big[P_\ell \bar J\big](s) &\equiv \frac{1}{2}\int_{-1}^1 \diff z P_\ell(z) \bar J\big(t(s,z)\big) \,, \notag \\
    t(s,z) &= \frac{1}{2}\big(4M^2-s\big)(1-z) \,,
\end{align}
which can easily be performed analytically.  The imaginary parts of the first few of these are given by
\begin{align}
    \Im \big[P_0 \bar J\big](s) &= - \frac{\pi}{\sigma} \big[1-(\sigma^2-1) L_\sigma \big]\theta(-s) \,,\\
    \Im \big[P_1 \bar J\big](s) &= \frac{\pi(1-\sigma^2)}{2\sigma^3} \big[1-(\sigma^2+1) L_\sigma \big]\theta(-s) \,, \notag\\
     \Im \big[P_2 \bar J\big](s) &= \frac{\pi(1-\sigma^2)}{4\sigma^5} \big[\sigma^2+3-(\sigma^4+3) L_\sigma \big]\theta(-s) \,.\notag
\end{align}
Near $s=0$, they have the common expansion
\begin{align}
    \Im \big[P_\ell \bar J\big](s) &= (-1)^{\ell+1} \frac{2\pi}{3\sigma^3} + \Order\big(\sigma^{-5}\big) \\
    &= (-1)^{\ell+1}\frac{\pi}{12M^3}(-s)^{3/2} + \Order\big((-s)^{5/2}\big) \,.\notag
\end{align}
This can be understood from the fact that the integration for the partial-wave projection hits the LHC first for $z = -1$, where $P_\ell(-1) = (-1)^\ell$.  We therefore conclude that left-hand singularities in $\pi\pi$ partial waves are of degree $(-s)^{3/2}$ (only).  This is reflected in the form of our conformal parameterization of the background amplitude in Sec.~\ref{sec:Swaves}.

\section{Isoscalar $S$-wave Adler zero}
\label{app:Adler}

In this appendix, we collect the higher-order corrections to the Adler zero in the $\pi\pi$ isospin-$0$ $S$-wave, $t^0_0(s)$. While the two-loop amplitude has been known since Refs.~\cite{Bijnens:1995yn,Bijnens:1997vq}, to the best of our knowledge, even the one-loop corrections to the LO Adler zero
\beq
s_A^{(2)}=\frac{\mpi^2}{2}
\eeq
have not been spelled out explicitly in the literature despite being used in, e.g., the modified inverse amplitude method~\cite{GomezNicola:2007qj}. Using $t^0_0(s)$ in the form given in Ref.~\cite{Niehus:2020gmf}, we find
\beq
s_A=s_A^{(2)}+s_A^{(4)}+s_A^{(6)}\, ,
\eeq
with 
\begin{align}
\label{Adler_zero}
 s_A^{(4)}&=-\frac{\mpi^4}{(48\pi F)^2}\Big[1163+2\big(107 \bar l_1+158\bar l_2-90 \bar l_3\big)\notag\\
 &\quad-908 A-4224 A^2\Big]\, ,\notag\\
 s_A^{(6)}&=-\frac{\mpi^6}{(12\pi F)^4}\bigg\{\frac{5}{64}\Big(\frac{17948821}{336} + 895 \pi^2\Big)\notag\\
 &
 -9A \Big(\frac{349339}{448}+20\pi^2\Big)\notag\\
&-\frac{A^2}{7}\Big[\frac{375125}{4}+1008\pi^2-34776\log\frac{7}{2}\notag\\
&\quad-6A\big(19073+1296 A\big)\Big]\notag\\
&+\frac{4968}{7}\Big[\text{Cl}_3\big(2\sqrt{7} A\big)+2\sqrt{7} A\, \text{Cl}_2\big(2\sqrt{7} A\big) - \zeta(3)\Big]\notag\\
&+\frac{1}{16}\Big[3103\bar l_1^2+7364\bar l_1\bar l_2+4108\bar l_2^2\notag\\
&\quad-2610 \bar l_1\bar l_3-2340\bar l_2\bar l_3-1620\bar l_3^2\Big]\notag\\
&+\frac{1}{896}\Big[
\bar l_1\big(1454303-1390328 A-4098432 A^2\big)\notag\\
&\quad+ \bar l_2\big(2410997-2315912 A-6820608 A^2\big) \notag\\
&\quad-1440\bar l_3\big(470-557A-984 A^2\big)\Big] \notag\\
& -\frac{L}{128}\Big(21653-12\big(17\bar l_1-5257\bar l_2+1080\bar l_3\big)\Big)\notag\\
&- 195 L^2+\frac{9}{32}(4\pi)^4 r \bigg\}\, ,
\end{align}
where 
\beq
A=\frac{\arctan\sqrt{7}}{\sqrt{7}}\, ,\qquad L=\log\frac{\mpi^2}{\mu^2}\, ,
\eeq
and the Clausen functions are related to polylogarithms via
\beq
\text{Cl}_3(\theta)=\Re\text{Li}_3\big(e^{i\theta}\big)\, , \quad 
\text{Cl}_2(\theta)=\Im\text{Li}_2\big(e^{i\theta}\big)\, .
\eeq
The $\bar l_i$ are given in the conventions of Ref.~\cite{Gasser:1983yg}, $\mpi$ is the physical pion mass, and $F$ the pion decay constant in the chiral limit. The two-loop low-energy constants are collected in 
\begin{align}
r&=3\big[240 (r_1^r+r_2^r)+428 r_3^r+836 r_4^r+1047 r_5^r+483 r_6^r\big]\notag\\
&\simeq -\frac{6F_\pi^2(3840 f_\chi^2+896\sqrt{2} f_\chi g_V+147 g_V^2)}{M_V^2}\notag\\
&\simeq -0.11\, , \label{eq:r-ressat}
\end{align}
where we used the resonance-saturation estimate from Ref.~\cite{Bijnens:1997vq} at $\mu=0.77\GeV$, with the phenomenological input from Ref.~\cite{Colangelo:2021moe}. We neglect uncertainties for the estimate Eq.~\eqref{eq:r-ressat}, since the impact of $r$ on $s_A^{(6)}$ already proves minor, especially in comparison to the dominant uncertainty due to $\bar l_1$ in $s_A^{(4)}$.
Based on the $\beta$ functions provided in Ref.~\cite{Bijnens:1999hw}, we also checked that the scale dependence of $r$ indeed cancels the one from the $L$ and $L^2$ terms in Eq.~\eqref{Adler_zero}. Numerically, we obtain 
\beq
\frac{s_A^{(4)}}{s_A^{(2)}}=-0.12(3)\, ,\qquad
\frac{s_A^{(6)}}{s_A^{(2)}}=-0.03(1)\, ,
\eeq
so that in combination $s_A$ reduces by $15(4)\%$ compared to its tree-level value. For the low-energy constants, we used the input $\bar l_1=-0.4(6)$, $\bar l_2=4.3(1)$~\cite{Colangelo:2001df}, $\bar l_3=3.4(3)$, and $F_\pi/F=1.07(1)$~\cite{FlavourLatticeAveragingGroupFLAG:2021npn,MILC:2010hzw,ETM:2010cqp,Beane:2011zm,Borsanyi:2012zv,BMW:2013fzj,Boyle:2015exm}.

\bibliographystyle{utphysmod.bst}
\balance
\bibliography{paper_refs}

\providecommand{\href}[2]{#2}\begingroup\raggedright\begin{thebibliography}{10}

\bibitem{Nakano:1982bc}
K.~Nakano, \href{http://dx.doi.org/10.1103/PhysRevC.26.1123}{Phys. Rev. C
  {\bfseries 26}, 1123 (1982)}.

\bibitem{Breit:1936zzb}
G.~Breit and E.~Wigner, \href{http://dx.doi.org/10.1103/PhysRev.49.519}{Phys.
  Rev. {\bfseries 49}, 519 (1936)}.

\bibitem{Gounaris:1968mw}
G.~J. Gounaris and J.~J. Sakurai,
  \href{http://dx.doi.org/10.1103/PhysRevLett.21.244}{Phys. Rev. Lett.
  {\bfseries 21}, 244 (1968)}.

\bibitem{Ananthanarayan:2000ht}
B.~Ananthanarayan, G.~Colangelo, J.~Gasser, and H.~Leutwyler,
  \href{http://dx.doi.org/10.1016/S0370-1573(01)00009-6}{Phys. Rept. {\bfseries
  353}, 207 (2001)}
[\href{https://arxiv.org/abs/hep-ph/0005297}{{arXiv:hep-ph/0005297}}].

\bibitem{Colangelo:2001df}
G.~Colangelo, J.~Gasser, and H.~Leutwyler,
  \href{http://dx.doi.org/10.1016/S0550-3213(01)00147-X}{Nucl. Phys. B
  {\bfseries 603}, 125 (2001)}
  [\href{https://arxiv.org/abs/hep-ph/0103088}{{arXiv:hep-ph/0103088}}].

\bibitem{Garcia-Martin:2011iqs}
R.~Garc\'ia-Mart\'in, R.~Kami\'nski, J.~R. Pel\'aez, J.~Ruiz~de Elvira, and
  F.~J. Yndur\'ain, \href{http://dx.doi.org/10.1103/PhysRevD.83.074004}{Phys.
  Rev. D {\bfseries 83}, 074004 (2011)}
  [\href{https://arxiv.org/abs/1102.2183}{{arXiv:1102.2183 [hep-ph]}}].

\bibitem{Caprini:2011ky}
I.~Caprini, G.~Colangelo, and H.~Leutwyler,
  \href{http://dx.doi.org/10.1140/epjc/s10052-012-1860-1}{Eur. Phys. J. C
  {\bfseries 72}, 1860 (2012)}
  [\href{https://arxiv.org/abs/1111.7160}{{arXiv:1111.7160 [hep-ph]}}].

\bibitem{Colangelo:2018mtw}
G.~Colangelo, M.~Hoferichter, and P.~Stoffer,
  \href{http://dx.doi.org/10.1007/JHEP02(2019)006}{JHEP {\bfseries 02}, 006
  (2019)} [\href{https://arxiv.org/abs/1810.00007}{{arXiv:1810.00007
  [hep-ph]}}].

\bibitem{Roy:1971tc}
S.~M. Roy, \href{http://dx.doi.org/10.1016/0370-2693(71)90724-6}{Phys. Lett. B
  {\bfseries 36}, 353 (1971)}.

\bibitem{Hanhart:2012wi}
C.~Hanhart, \href{http://dx.doi.org/10.1016/j.physletb.2012.07.038}{Phys. Lett.
  B {\bfseries 715}, 170 (2012)}
  [\href{https://arxiv.org/abs/1203.6839}{{arXiv:1203.6839 [hep-ph]}}].

\bibitem{Chanturia:2022rcz}
G.~Chanturia, \href{http://dx.doi.org/10.22323/1.412.0030}{PoS {\bfseries
  Regio2021}, 030 (2022)}.

\bibitem{Colangelo:2020lcg}
G.~Colangelo, M.~Hoferichter, and P.~Stoffer,
  \href{http://dx.doi.org/10.1016/j.physletb.2021.136073}{Phys. Lett. B
  {\bfseries 814}, 136073 (2021)}
  [\href{https://arxiv.org/abs/2010.07943}{{arXiv:2010.07943 [hep-ph]}}].

\bibitem{Colangelo:2022prz}
G.~Colangelo, M.~Hoferichter, B.~Kubis, and P.~Stoffer,
  \href{http://dx.doi.org/10.1007/JHEP10(2022)032}{JHEP {\bfseries 10}, 032
  (2022)} [\href{https://arxiv.org/abs/2208.08993}{{arXiv:2208.08993
  [hep-ph]}}].

\bibitem{Pelaez:2015qba}
J.~R. Pel\'aez, \href{http://dx.doi.org/10.1016/j.physrep.2016.09.001}{Phys.
  Rept. {\bfseries 658}, 1 (2016)}
  [\href{https://arxiv.org/abs/1510.00653}{{arXiv:1510.00653 [hep-ph]}}].

\bibitem{Caprini:2005zr}
I.~Caprini, G.~Colangelo, and H.~Leutwyler,
  \href{http://dx.doi.org/10.1103/PhysRevLett.96.132001}{Phys. Rev. Lett.
  {\bfseries 96}, 132001 (2006)}
  [\href{https://arxiv.org/abs/hep-ph/0512364}{{arXiv:hep-ph/0512364}}].

\bibitem{Garcia-Martin:2011nna}
R.~Garc\'ia-Mart\'in, R.~Kami\'nski, J.~R. Pel\'aez, and J.~Ruiz~de Elvira,
  \href{http://dx.doi.org/10.1103/PhysRevLett.107.072001}{Phys. Rev. Lett.
  {\bfseries 107}, 072001 (2011)}
  [\href{https://arxiv.org/abs/1107.1635}{{arXiv:1107.1635 [hep-ph]}}].

\bibitem{Moussallam:2011zg}
B.~Moussallam, \href{http://dx.doi.org/10.1140/epjc/s10052-011-1814-z}{Eur.
  Phys. J. C {\bfseries 71}, 1814 (2011)}
  [\href{https://arxiv.org/abs/1110.6074}{{arXiv:1110.6074 [hep-ph]}}].

\bibitem{Donoghue:1990xh}
J.~F. Donoghue, J.~Gasser, and H.~Leutwyler,
  \href{http://dx.doi.org/10.1016/0550-3213(90)90474-R}{Nucl. Phys. B
  {\bfseries 343}, 341 (1990)}.

\bibitem{Gardner:2001gc}
S.~Gardner and U.-G. Mei{\ss}ner,
  \href{http://dx.doi.org/10.1103/PhysRevD.65.094004}{Phys. Rev. D {\bfseries
  65}, 094004 (2002)}
  [\href{https://arxiv.org/abs/hep-ph/0112281}{{arXiv:hep-ph/0112281}}].

\bibitem{Hoferichter:2012wf}
M.~Hoferichter, C.~Ditsche, B.~Kubis, and U.-G. Mei{\ss}ner,
  \href{http://dx.doi.org/10.1007/JHEP06(2012)063}{JHEP {\bfseries 06}, 063
  (2012)} [\href{https://arxiv.org/abs/1204.6251}{{arXiv:1204.6251 [hep-ph]}}].

\bibitem{Daub:2015xja}
J.~T. Daub, C.~Hanhart, and B.~Kubis,
  \href{http://dx.doi.org/10.1007/JHEP02(2016)009}{JHEP {\bfseries 02}, 009
  (2016)} [\href{https://arxiv.org/abs/1508.06841}{{arXiv:1508.06841
  [hep-ph]}}].

\bibitem{Adler:1964um}
S.~L. Adler, \href{http://dx.doi.org/10.1103/PhysRev.137.B1022}{Phys. Rev.
  {\bfseries 137}, B1022 (1965)}.

\bibitem{Adler:1965ga}
S.~L. Adler, \href{http://dx.doi.org/10.1103/PhysRev.139.B1638}{Phys. Rev.
  {\bfseries 139}, B1638 (1965)}.

\bibitem{Dobado:1996ps}
A.~Dobado and J.~R. Pel\'aez,
  \href{http://dx.doi.org/10.1103/PhysRevD.56.3057}{Phys. Rev. D {\bfseries
  56}, 3057 (1997)}
  [\href{https://arxiv.org/abs/hep-ph/9604416}{{arXiv:hep-ph/9604416}}].

\bibitem{GomezNicola:2007qj}
A.~G{\'o}mez~Nicola, J.~R. Pel{\'a}ez, and G.~R{\'i}os,
  \href{http://dx.doi.org/10.1103/PhysRevD.77.056006}{Phys. Rev. D {\bfseries
  77}, 056006 (2008)} [\href{https://arxiv.org/abs/0712.2763}{{arXiv:0712.2763
  [hep-ph]}}].

\bibitem{Wang:2022vga}
Z.-Q. Wang, X.-W. Kang, J.~A. Oller, and L.~Zhang,
  \href{http://dx.doi.org/10.1103/PhysRevD.105.074016}{Phys. Rev. D {\bfseries
  105}, 074016 (2022)}
  [\href{https://arxiv.org/abs/2201.00492}{{arXiv:2201.00492 [hep-ph]}}].

\bibitem{Svarc:2022buh}
A.~\v{S}varc and R.~L. Workman,
  \href{http://dx.doi.org/10.1103/PhysRevC.108.014615}{Phys. Rev. C {\bfseries
  108}, 014615 (2023)}
  [\href{https://arxiv.org/abs/2206.05978}{{arXiv:2206.05978 [nucl-th]}}].

\bibitem{Burkert:2022bqo}
V.~Burkert {\em et~al.},
  \href{http://dx.doi.org/10.1016/j.physletb.2023.138070}{Phys. Lett. B
  {\bfseries 844}, 138070 (2023)}
  [\href{https://arxiv.org/abs/2207.08472}{{arXiv:2207.08472 [hep-ph]}}].

\bibitem{Battaglieri:2014gca}
M.~Battaglieri {\em et~al.},
  \href{http://dx.doi.org/10.5506/APhysPolB.46.257}{Acta Phys. Polon. B
  {\bfseries 46}, 257 (2015)}
  [\href{https://arxiv.org/abs/1412.6393}{{arXiv:1412.6393 [hep-ph]}}].

\bibitem{Morgan:1987gv}
D.~Morgan and M.~R. Pennington, \href{http://dx.doi.org/10.1007/BF01578139}{Z.
  Phys. {\bfseries C37}, 431 (1988)},
[Erratum: Z. Phys. \textbf{C39}, 590 (1988)].

\bibitem{Morgan:1990kw}
D.~Morgan and M.~R. Pennington,
\href{http://dx.doi.org/10.1007/BF01614697}{Z. Phys. {\bfseries C48}, 623
  (1990)}.

\bibitem{Hoferichter:2011wk}
M.~Hoferichter, D.~R. Phillips, and C.~Schat,
  \href{http://dx.doi.org/10.1140/epjc/s10052-011-1743-x}{Eur. Phys. J. C
  {\bfseries 71}, 1743 (2011)}
  [\href{https://arxiv.org/abs/1106.4147}{{arXiv:1106.4147 [hep-ph]}}].

\bibitem{Danilkin:2020pak}
I.~Danilkin, O.~Deineka, and M.~Vanderhaeghen,
  \href{http://dx.doi.org/10.1103/PhysRevD.103.114023}{Phys. Rev. D {\bfseries
  103}, 114023 (2021)}
  [\href{https://arxiv.org/abs/2012.11636}{{arXiv:2012.11636 [hep-ph]}}].

\bibitem{Zou:1993az}
B.~S. Zou and D.~V. Bugg,
  \href{http://dx.doi.org/10.1103/PhysRevD.48.R3948}{Phys. Rev. D {\bfseries
  48}, R3948 (1993)}.

\bibitem{Albaladejo:2015aca}
M.~Albaladejo and B.~Moussallam,
  \href{http://dx.doi.org/10.1140/epjc/s10052-015-3715-z}{Eur. Phys. J. C
  {\bfseries 75}, 488 (2015)}
  [\href{https://arxiv.org/abs/1507.04526}{{arXiv:1507.04526 [hep-ph]}}].

\bibitem{Lu:2020qeo}
J.~Lu and B.~Moussallam,
  \href{http://dx.doi.org/10.1140/epjc/s10052-020-7969-8}{Eur. Phys. J. C
  {\bfseries 80}, 436 (2020)}
  [\href{https://arxiv.org/abs/2002.04441}{{arXiv:2002.04441 [hep-ph]}}].

\bibitem{Ropertz:2018stk}
S.~Ropertz, C.~Hanhart, and B.~Kubis,
  \href{http://dx.doi.org/10.1140/epjc/s10052-018-6416-6}{Eur. Phys. J. C
  {\bfseries 78}, 1000 (2018)}
  [\href{https://arxiv.org/abs/1809.06867}{{arXiv:1809.06867 [hep-ph]}}].

\bibitem{VonDetten:2021rax}
L.~von Detten, F.~No\"el, C.~Hanhart, M.~Hoferichter, and B.~Kubis,
  \href{http://dx.doi.org/10.1140/epjc/s10052-021-09169-7}{Eur. Phys. J. C
  {\bfseries 81}, 420 (2021)}
  [\href{https://arxiv.org/abs/2103.01966}{{arXiv:2103.01966 [hep-ph]}}].

\bibitem{Gasparyan:2012km}
A.~M. Gasparyan, M.~F.~M. Lutz, and E.~Epelbaum,
  \href{http://dx.doi.org/10.1140/epja/i2013-13115-7}{Eur. Phys. J. A
  {\bfseries 49}, 115 (2013)}
  [\href{https://arxiv.org/abs/1212.3057}{{arXiv:1212.3057 [nucl-th]}}].

\bibitem{Pelaez:2019eqa}
J.~R. Pel\'aez, A.~Rodas, and J.~Ruiz~de Elvira,
  \href{http://dx.doi.org/10.1140/epjc/s10052-019-7509-6}{Eur. Phys. J. C
  {\bfseries 79}, 1008 (2019)}
  [\href{https://arxiv.org/abs/1907.13162}{{arXiv:1907.13162 [hep-ph]}}].

\bibitem{Danilkin:2022cnj}
I.~Danilkin, V.~Biloshytskyi, X.-L. Ren, and M.~Vanderhaeghen,
  \href{http://dx.doi.org/10.1103/PhysRevD.107.074021}{Phys. Rev. D {\bfseries
  107}, 074021 (2023)}
  [\href{https://arxiv.org/abs/2206.15223}{{arXiv:2206.15223 [hep-ph]}}].

\bibitem{Omnes:1958hv}
R.~Omn\`es, \href{http://dx.doi.org/10.1007/BF02747746}{Nuovo Cim. {\bfseries
  8}, 316 (1958)}.

\bibitem{Weinberg:1995mt}
S.~Weinberg, {\em {The Quantum theory of fields. Vol. 1: Foundations}},
  Cambridge University Press, 6, 2005.

\bibitem{VonHippel:1972fg}
F.~von Hippel and C.~Quigg,
  \href{http://dx.doi.org/10.1103/PhysRevD.5.624}{Phys. Rev. D {\bfseries 5},
  624 (1972)}.

\bibitem{Chung:1995dx}
S.~U. Chung, J.~Brose, R.~Hackmann, E.~Klempt, S.~Spanier, and C.~Strassburger,
  \href{http://dx.doi.org/10.1002/andp.19955070504}{Annalen Phys. {\bfseries
  4}, 404 (1995)}.

\bibitem{Lomon:2012pn}
E.~L. Lomon and S.~Pacetti,
  \href{http://dx.doi.org/10.1103/PhysRevD.86.039901}{Phys. Rev. D {\bfseries
  85}, 113004 (2012)} [\href{https://arxiv.org/abs/1201.6126}{{arXiv:1201.6126
  [hep-ph]}}], [Erratum: Phys. Rev. D {\bf 86}, 039901 (2012)].

\bibitem{Moussallam:2013una}
B.~Moussallam, \href{http://dx.doi.org/10.1140/epjc/s10052-013-2539-y}{Eur.
  Phys. J. C {\bfseries 73}, 2539 (2013)}
  [\href{https://arxiv.org/abs/1305.3143}{{arXiv:1305.3143 [hep-ph]}}].

\bibitem{Zanke:2021wiq}
M.~Zanke, M.~Hoferichter, and B.~Kubis,
  \href{http://dx.doi.org/10.1007/JHEP07(2021)106}{JHEP {\bfseries 07}, 106
  (2021)} [\href{https://arxiv.org/abs/2103.09829}{{arXiv:2103.09829
  [hep-ph]}}].

\bibitem{Crivellin:2022gfu}
A.~Crivellin and M.~Hoferichter,
  \href{http://dx.doi.org/10.1103/PhysRevD.108.013005}{Phys. Rev. D {\bfseries
  108}, 013005 (2023)}
  [\href{https://arxiv.org/abs/2211.12516}{{arXiv:2211.12516 [hep-ph]}}].

\bibitem{Kallen:1952zz}
G.~K{\"a}ll{\'e}n, \href{http://dx.doi.org/10.1007/978-3-319-00627-7_90}{Helv.
  Phys. Acta {\bfseries 25}, 417 (1952)}.

\bibitem{Lehmann:1954xi}
H.~Lehmann, \href{http://dx.doi.org/10.1007/BF02783624}{Nuovo Cim. {\bfseries
  11}, 342 (1954)}.

\bibitem{Hoferichter:2023mgy}
M.~Hoferichter, J.~Ruiz~de Elvira, B.~Kubis, and U.-G. Mei\ss{}ner,
  \href{http://dx.doi.org/10.1016/j.physletb.2024.138698}{Phys. Lett. B
  {\bfseries 853}, 138698 (2024)}
  [\href{https://arxiv.org/abs/2312.15015}{{arXiv:2312.15015 [hep-ph]}}].

\bibitem{Belle:2008xpe}
M.~Fujikawa {\em et~al.} [Belle Collaboration],
  \href{http://dx.doi.org/10.1103/PhysRevD.78.072006}{Phys. Rev. D {\bfseries
  78}, 072006 (2008)} [\href{https://arxiv.org/abs/0805.3773}{{arXiv:0805.3773
  [hep-ex]}}].

\bibitem{Chernyak:1977as}
V.~L. Chernyak and A.~R. Zhitnitsky, JETP Lett. {\bfseries 25}, 510 (1977),
  [Zh. Eksp. Teor. Fiz. {\bf 25}, 544 (1977)].

\bibitem{Farrar:1979aw}
G.~R. Farrar and D.~R. Jackson,
  \href{http://dx.doi.org/10.1103/PhysRevLett.43.246}{Phys. Rev. Lett.
  {\bfseries 43}, 246 (1979)}.

\bibitem{Efremov:1979qk}
A.~V. Efremov and A.~V. Radyushkin,
  \href{http://dx.doi.org/10.1016/0370-2693(80)90869-2}{Phys. Lett. B
  {\bfseries 94}, 245 (1980)}.

\bibitem{Lepage:1979zb}
G.~P. Lepage and S.~J. Brodsky,
  \href{http://dx.doi.org/10.1016/0370-2693(79)90554-9}{Phys. Lett. B
  {\bfseries 87}, 359 (1979)}.

\bibitem{Lepage:1980fj}
G.~P. Lepage and S.~J. Brodsky,
  \href{http://dx.doi.org/10.1103/PhysRevD.22.2157}{Phys. Rev. D {\bfseries
  22}, 2157 (1980)}.

\bibitem{NA482:2010dug}
J.~R. Batley {\em et~al.} [NA48/2 Collaboration],
  \href{http://dx.doi.org/10.1140/epjc/s10052-010-1480-6}{Eur. Phys. J. C
  {\bfseries 70}, 635 (2010)}.

\bibitem{Xiao:2000kx}
Z.~Xiao and H.~Q. Zheng,
  \href{http://dx.doi.org/10.1016/S0375-9474(01)01100-9}{Nucl. Phys. A
  {\bfseries 695}, 273 (2001)}
  [\href{https://arxiv.org/abs/hep-ph/0011260}{{arXiv:hep-ph/0011260}}].

\bibitem{Willenbrock:2022smq}
S.~Willenbrock [\href{https://arxiv.org/abs/2203.11056}{{arXiv:2203.11056
  [hep-ph]}}].

\bibitem{Hoferichter:2017ftn}
M.~Hoferichter, B.~Kubis, and M.~Zanke,
  \href{http://dx.doi.org/10.1103/PhysRevD.96.114016}{Phys. Rev. D {\bfseries
  96}, 114016 (2017)}
  [\href{https://arxiv.org/abs/1710.00824}{{arXiv:1710.00824 [hep-ph]}}].

\bibitem{Meissner:1987ge}
U.-G. Mei{\ss}ner, \href{http://dx.doi.org/10.1016/0370-1573(88)90090-7}{Phys.
  Rept. {\bfseries 161}, 213 (1988)}.

\bibitem{Klingl:1996by}
F.~Klingl, N.~Kaiser, and W.~Weise,
  \href{http://dx.doi.org/10.1007/s002180050167}{Z. Phys. A {\bfseries 356},
  193 (1996)}
  [\href{https://arxiv.org/abs/hep-ph/9607431}{{arXiv:hep-ph/9607431}}].

\bibitem{Stamen:2022uqh}
D.~Stamen, D.~Hariharan, M.~Hoferichter, B.~Kubis, and P.~Stoffer,
  \href{http://dx.doi.org/10.1140/epjc/s10052-022-10348-3}{Eur. Phys. J. C
  {\bfseries 82}, 432 (2022)}
  [\href{https://arxiv.org/abs/2202.11106}{{arXiv:2202.11106 [hep-ph]}}].

\bibitem{Lees:2012cj}
J.~P. Lees {\em et~al.} [BaBar Collaboration],
  \href{http://dx.doi.org/10.1103/PhysRevD.86.032013}{Phys. Rev. D {\bfseries
  86}, 032013 (2012)} [\href{https://arxiv.org/abs/1205.2228}{{arXiv:1205.2228
  [hep-ex]}}].

\bibitem{Anastasi:2017eio}
A.~Anastasi {\em et~al.} [KLOE-2 Collaboration],
  \href{http://dx.doi.org/10.1007/JHEP03(2018)173}{JHEP {\bfseries 03}, 173
  (2018)} [\href{https://arxiv.org/abs/1711.03085}{{arXiv:1711.03085
  [hep-ex]}}].

\bibitem{CMD-3:2023alj}
F.~V. Ignatov {\em et~al.} [CMD-3 Collaboration]
  [\href{https://arxiv.org/abs/2302.08834}{{arXiv:2302.08834 [hep-ex]}}].

\bibitem{Colangelo:2023rqr}
G.~Colangelo, M.~Hoferichter, and P.~Stoffer
  [\href{https://arxiv.org/abs/2308.04217}{{arXiv:2308.04217 [hep-ph]}}].

\bibitem{Bijnens:1995yn}
J.~Bijnens, G.~Colangelo, G.~Ecker, J.~Gasser, and M.~E. Sainio,
  \href{http://dx.doi.org/10.1016/0370-2693(96)00165-7}{Phys. Lett. B
  {\bfseries 374}, 210 (1996)}
  [\href{https://arxiv.org/abs/hep-ph/9511397}{{arXiv:hep-ph/9511397}}].

\bibitem{Bijnens:1997vq}
J.~Bijnens, G.~Colangelo, G.~Ecker, J.~Gasser, and M.~E. Sainio,
  \href{http://dx.doi.org/10.1016/S0550-3213(97)00621-4}{Nucl. Phys. B
  {\bfseries 508}, 263 (1997)}
  [\href{https://arxiv.org/abs/hep-ph/9707291}{{arXiv:hep-ph/9707291}}],
  [Erratum: Nucl. Phys. B {\bf 517}, 639 (1998)].

\bibitem{Niehus:2020gmf}
M.~Niehus, M.~Hoferichter, B.~Kubis, and J.~Ruiz~de Elvira,
  \href{http://dx.doi.org/10.1103/PhysRevLett.126.102002}{Phys. Rev. Lett.
  {\bfseries 126}, 102002 (2021)}
  [\href{https://arxiv.org/abs/2009.04479}{{arXiv:2009.04479 [hep-ph]}}].

\bibitem{Gasser:1983yg}
J.~Gasser and H.~Leutwyler,
  \href{http://dx.doi.org/10.1016/0003-4916(84)90242-2}{Annals Phys. {\bfseries
  158}, 142 (1984)}.

\bibitem{Colangelo:2021moe}
G.~Colangelo, M.~Hoferichter, B.~Kubis, M.~Niehus, and J.~Ruiz~de Elvira,
  \href{http://dx.doi.org/10.1016/j.physletb.2021.136852}{Phys. Lett. B
  {\bfseries 825}, 136852 (2022)}
  [\href{https://arxiv.org/abs/2110.05493}{{arXiv:2110.05493 [hep-ph]}}].

\bibitem{Bijnens:1999hw}
J.~Bijnens, G.~Colangelo, and G.~Ecker,
  \href{http://dx.doi.org/10.1006/aphy.1999.5982}{Annals Phys. {\bfseries 280},
  100 (2000)}
  [\href{https://arxiv.org/abs/hep-ph/9907333}{{arXiv:hep-ph/9907333}}].

\bibitem{FlavourLatticeAveragingGroupFLAG:2021npn}
Y.~Aoki {\em et~al.} [Flavour Lattice Averaging Group (FLAG) Collaboration],
  \href{http://dx.doi.org/10.1140/epjc/s10052-022-10536-1}{Eur. Phys. J. C
  {\bfseries 82}, 869 (2022)}
  [\href{https://arxiv.org/abs/2111.09849}{{arXiv:2111.09849 [hep-lat]}}].

\bibitem{MILC:2010hzw}
A.~Bazavov {\em et~al.} [MILC Collaboration],
  \href{http://dx.doi.org/10.22323/1.105.0074}{PoS {\bfseries LATTICE2010}, 074
  (2010)} [\href{https://arxiv.org/abs/1012.0868}{{arXiv:1012.0868
  [hep-lat]}}].

\bibitem{ETM:2010cqp}
R.~Baron {\em et~al.} [ETM Collaboration],
  \href{http://dx.doi.org/10.22323/1.105.0123}{PoS {\bfseries LATTICE2010}, 123
  (2010)} [\href{https://arxiv.org/abs/1101.0518}{{arXiv:1101.0518
  [hep-lat]}}].

\bibitem{Beane:2011zm}
S.~R. Beane {\em et~al.} [NPLQCD Collaboration],
  \href{http://dx.doi.org/10.1103/PhysRevD.86.094509}{Phys. Rev. D {\bfseries
  86}, 094509 (2012)} [\href{https://arxiv.org/abs/1108.1380}{{arXiv:1108.1380
  [hep-lat]}}].

\bibitem{Borsanyi:2012zv}
S.~Borsanyi {\em et~al.} [BMW Collaboration],
  \href{http://dx.doi.org/10.1103/PhysRevD.88.014513}{Phys. Rev. D {\bfseries
  88}, 014513 (2013)} [\href{https://arxiv.org/abs/1205.0788}{{arXiv:1205.0788
  [hep-lat]}}].

\bibitem{BMW:2013fzj}
S.~D\"urr {\em et~al.} [BMW Collaboration],
  \href{http://dx.doi.org/10.1103/PhysRevD.90.114504}{Phys. Rev. D {\bfseries
  90}, 114504 (2014)} [\href{https://arxiv.org/abs/1310.3626}{{arXiv:1310.3626
  [hep-lat]}}].

\bibitem{Boyle:2015exm}
P.~A. Boyle {\em et~al.} [RBC/UKQCD Collaboration],
  \href{http://dx.doi.org/10.1103/PhysRevD.93.054502}{Phys. Rev. D {\bfseries
  93}, 054502 (2016)}
  [\href{https://arxiv.org/abs/1511.01950}{{arXiv:1511.01950 [hep-lat]}}].

\end{thebibliography}\endgroup

\end{document}